\documentclass[journal]{IEEEtran}

\usepackage{cite}
\usepackage{amsmath,amssymb,amsfonts}
\usepackage{algorithmic}
\usepackage{graphicx}
\usepackage{textcomp}
\usepackage{indentfirst}
\usepackage{color, colortbl}
\usepackage[table, svgnames, dvipsnames]{xcolor}
\usepackage{caption}
\usepackage[english]{babel}
\usepackage{wrapfig}
\usepackage{enumitem}
\usepackage{subcaption}
\usepackage{xfrac}
\usepackage{diagbox}
\usepackage{array}
\usepackage{multirow}

\newcolumntype{P}[1]{>{\centering\arraybackslash}p{#1}}

\ifCLASSINFOpdf
\else
\fi
\begin{document}

\title{Speed Advisory System Using Real-Time Actuated Traffic Light Phase Length Prediction}

\author{Mikhail~Burov,
        Murat~Arcak
        and~Alexander~Kurzhanskiy%
\thanks{Manuscript received ...}%
\thanks{M. Burov and M. Arcak are with the Department
of Electrical Engineering and Computer Sciences, University of California, Berkeley, CA, 94704 USA (e-mails: mikaburov@berkeley.edu; arcak@.berkeley.edu).}%
\thanks{A. Kurzhanskiy is with PATH, Berkeley, CA, 94720, USA (e-mail: akurzhan@berkeley.edu).}
\thanks{This project was sponsored by the NSF CPS grant CNS-1545116 - Traffic Operating System for Smart Cities.}}

\mark{TThis work has been submitted to the IEEE for possible publication. Copyright may be transferred without notice, after which this version may no longer be accessible.}

\maketitle

\begin{abstract}
Speed advisory systems for connected vehicles rely on the estimation of green (or red) light duration at signalized intersections. A particular challenge is to predict the signal phases of semi- and fully-actuated traffic lights.
In this paper, we introduce an algorithm that processes traffic measurement data collected from advanced detectors on road links and assigns ``PASS"/``WAIT" labels to connected vehicles according to their predicted ability to go through the upcoming signalized intersection within the current phase.
Additional computations provide an estimate for the duration of the current green phase that can be used by the Speed Advisory System to minimize fuel consumption.
Simulation results show 95\% prediction accuracy, which yields up to 30\% reduction in fuel consumption when used in a driver-assistance system. Traffic progression quality also benefits from our algorithm demonstrating an improvement of 20\% at peak for medium traffic demand, reducing delays and idling at intersections.
\end{abstract}

\begin{IEEEkeywords}
Speed Advisory System, connected vehicles, actuated traffic light, phase prediction, real-time algorithm
\end{IEEEkeywords}

%

\section{Introduction}

\IEEEPARstart{V}{ehicles} equipped with the Speed Advisory System (SAS) \cite{1} use traffic light information and traffic data to obtain the near-optimal speed trajectories to reduce fuel consumption. These trajectories also benefit progression quality by minimizing idling at intersections. One of the key parameters required by the SAS is the estimated remaining time until the end of the current traffic light (TL) phase. In the case of static traffic light (constant phase length) this information can be easily obtained directly from the traffic light via Signal Phase and Timing (SPaT) messages or any other signaling system. However, if an intersection is equipped with an actuated TL, the phase length depends on the traffic demand (vehicle flow) and, therefore, may vary from ``minimum duration" to ``maximum duration" - specific variables characterizing a particular traffic light. Since the exact phase length value is unavailable, a prediction algorithm is required to provide an estimation to be used in near-optimal speed trajectory derivation.  

The software presented in \cite{2} encourages drivers to use smartphone cameras to identify the traffic light color at the upcoming intersection and estimate the remaining time within the current phase. The study reports error rates from 7.8\% to 12.4\%. The phase length estimation, based on the five previous green-red/red-green transitions is also inefficient: according to the algorithm, the best prediction is just slightly better than estimating the current phase length to be the same as the previous phase length.

A follow-on study \cite{3} addresses the problem of finding optimal speed trajectories to minimize fuel consumption. However, the work considers only pre-timed signals, leaving behind issues with adaptive phase duration. The studies \cite{4} and \cite{5} also focus on pre-timed traffic lights.

References \cite{2} and \cite{6} use noisy measurements of a signal phase to process SPaT estimation. The study analyzes a large number of GPS position and speed samples from 4300 buses within a period of one month to estimate phase duration, cycle length and cycle start time. However, a large percentage of the recorded data was not suitable for the analysis, which limited the accuracy of the algorithm (6s error for 36s phase duration).
The later study \cite{7} estimates the waiting time spent by buses in queues and presents significantly better results for the SPaT estimate.
	
All of the noisy measurement-based algorithms are implemented only for pre-timed traffic lights. In addition, collecting and processing noisy measurements may be computationally inefficient, since most of the signal data are available from transportation authorities.

The study \cite{8} makes probabilistic SPaT predictions based on the intersection traffic data. At the beginning of each cycle, the empirical frequency distribution is computed. Furthermore, for every second within the cycle, the algorithm tries to predict whether a certain phase is \textbf{G}(green), \textbf{R}(red) or \textbf{M}(uncertain) with some level of confidence. Further estimation of the phase residual time requires the knowledge of within-cycle time, which even if available does not guarantee the accuracy of the prediction. In addition to the fact that ``80\% confidence" predictions may be incorrect, the algorithm does not provide firm guarantees and uncertainty may grow with the increase of confidence level.

An algorithm presented in \cite{9} relies on the historical data from several intersections in Munich and uses a Kalman Filter to predict future probability distributions of phase durations. Although a high level of accuracy was achieved (95\%), the practical applicability is limited since the availability level is only 71\% on average.

Another study \cite{10} suggests using both the historical phase measurements and the real-time information that locates the current time within the current phase to predict all future phase transitions. Two approaches are considered: ``conditional expectation based prediction" and ``confidence based prediction". These methods greatly improve the prediction of the residual time for the current phase as well as for the subsequent phase; however, as stated in the paper, the proposed algorithms ``pose a challenge to the design of speed profiles that reduce fuel consumption". Since both algorithms update their predictions every second, SAS-equipped vehicles are forced to reevaluate the speed trajectories every second as well, causing jerky motion and fuel consumption increase.
Several other papers study vehicle flow estimation at arterial roads using adaptive signal control predictions \cite{11}, \cite{12}, \cite{13}. 


Most of the algorithms discussed above try to estimate or predict the phase length/residual time of the phase using historical data and statistical methods. In contrast, our primary objective is to determine whether or not a vehicle can pass the upcoming intersection during the current green phase. Using real-time traffic data from advanced detectors and the upcoming actuated TL, the algorithm analyzes downstream traffic, estimates the time vehicles can reach the TL and assigns labels (``PASS"/ ``WAIT") depending on the vehicle's passing capability. According to the obtained labels, the algorithm derives an estimated phase residual time for near-optimal speed trajectory computation for every participating vehicle.

\section{The Algorithm}

We consider traffic lights with fixed cycle length and one actuated axis
(later in the paper ``green / red / yellow phase" refers to the phase of the actuated axis). 
Therefore, knowing the time within the cycle allows us to compute the precise remaining time until the next cycle (i.e. next green phase, assuming that every cycle starts with the green phase). In other words, for a non-green phase the remaining time until the next green is known and no predictions are required. However, if the current phase is green, the algorithm is used to estimate the phase length.

All the computations and simulations were conducted in an open source simulator SUMO (Simulation of Urban Mobility).

\subsection{Simple and Complex Network Architecture}

\begin{figure}[ht]
    \includegraphics[width=1.0\linewidth]{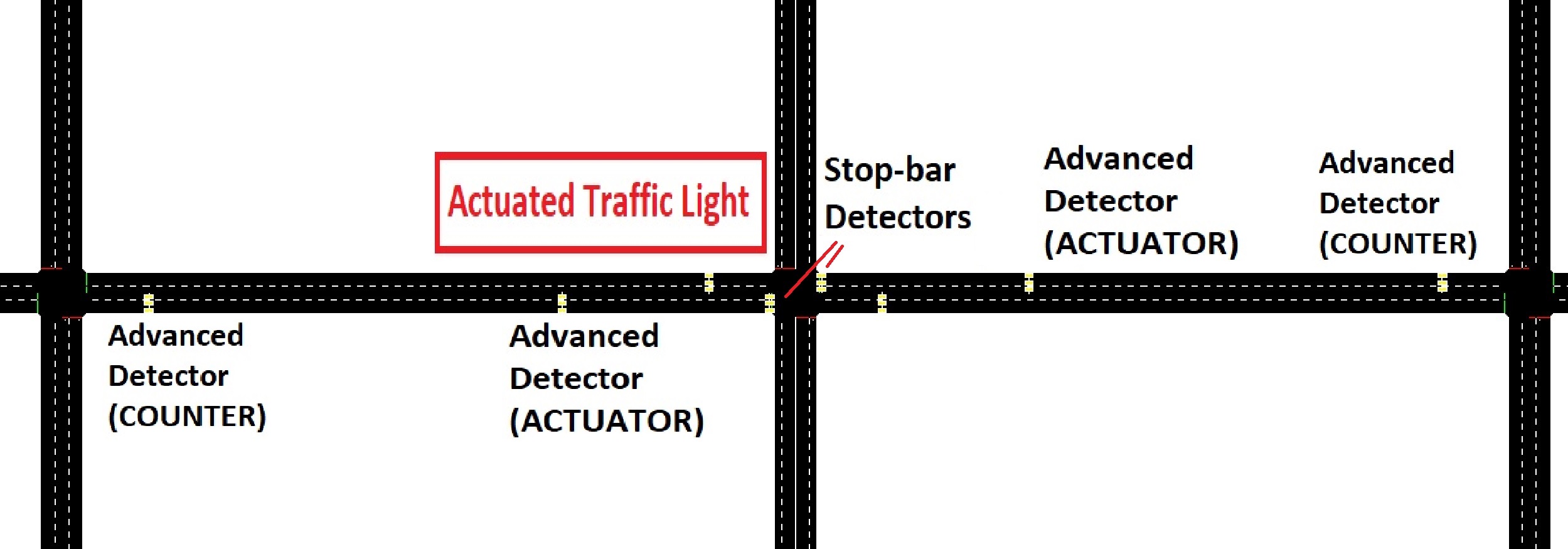}
  	\caption{Simple Network.}
  	\label{simpNet}
\end{figure}

\begin{figure}[ht]
  	\includegraphics[width=1.0\linewidth]{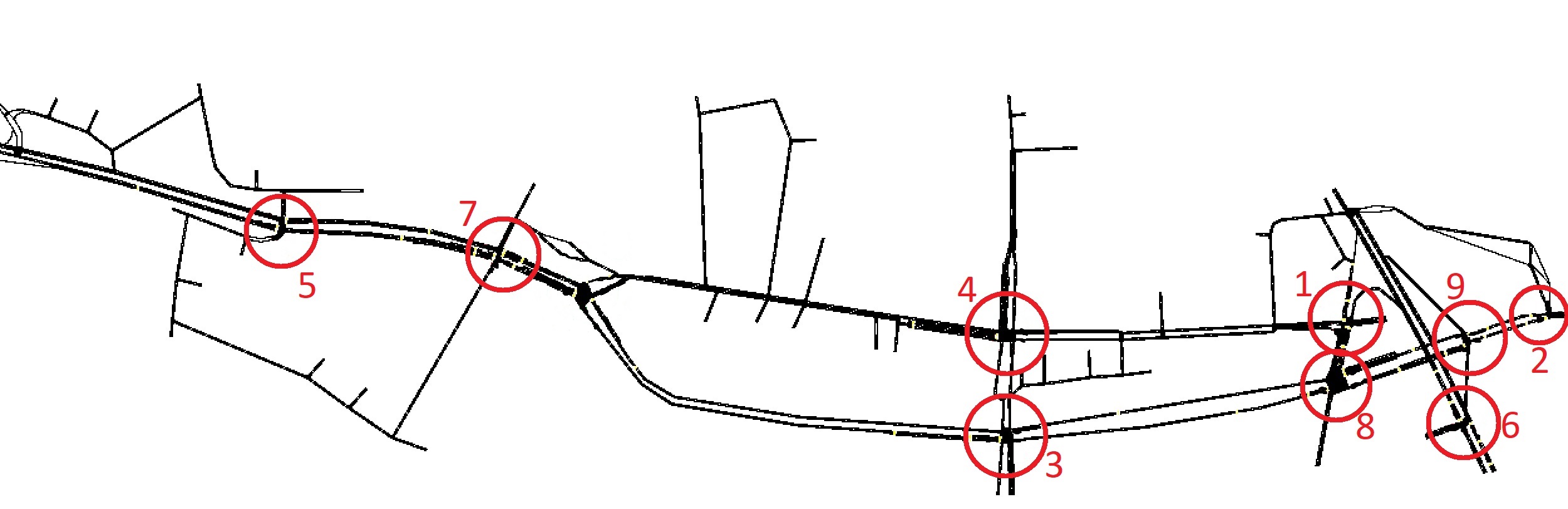}
  	\caption{North Bethesda, Montgomery County, MD.}
  	\label{compNet}
\end{figure}

The project consists of two parts. First, we analyze a simple symmetric signalized intersection (Fig. \ref{simpNet}) with East $\leftrightarrow$ West actuated axis to test an idealistic set-up and obtain a benchmark for further evaluation. The second part is a simulation of North Bethesda, Montgomery County, Maryland network (around the intersections of Montrose Rd and Montrose Pkwy), a complex system of 9 actuated traffic lights with different geometries and signal schedules (Fig. \ref{compNet}). This set-up allows us to test the algorithm in realistic conditions in the presence of various uncertainties.

Every incoming link on the actuated axis is equipped with a stop-bar detector and two advanced detectors: an ``actuator" and a ``counter". Actuators are responsible for prolonging the green phase when a vehicle is detected. If all necessary conditions are satisfied, a vehicle crossing an actuator triggers the green light extension. Actuators are placed 40-100 meters before the intersection. 

Counters are advanced detectors that collect necessary information about the traffic state and are located at least 50 meters upstream from actuators. When a vehicle crosses a counter, the infrastructure records and stores the vehicle's speed and time of passing for one cycle. At the end of every cycle, the data are erased, since it is no longer relevant to the current phase actuation. We assume $T_{th}$ seconds is enough to travel from the counter to the intersection in moderate traffic, therefore, counter data corresponding to the previous cycle has no impact on the current one. The case when the vehicle fails to reach the intersection within $T_{th}$ seconds implies congestion that forces the vehicle to switch to the car-following model.

\subsection{Traffic Light Properties}

As stated earlier, the cycle length of the TL $j$ is fixed and equals $cycLen_{j} \in [90, 120]$ seconds. The cycle consists of 4 phases in the simple case (Tab. \ref{tlParam})  and up to 8 phases in the complex set-up.

\begin{table}[htbp]
    \caption{Traffic light states: groups of three from left to right - North $\rightarrow$ South, West $\rightarrow$ East, South $\rightarrow$ North, East $\rightarrow$ West; r - red, G - green, y - yellow}
	\center
	\begin{tabular}{ |p{0.8cm}|p{2cm}|p{1.7cm}|p{1.7cm}|  }
 	    \hline
 	    Phase & TL States & Min Duration & Max Duration\\
 		\hline \hline
		\rowcolor{Gainsboro!60}
 		0 & rrrGGGrrrGGG  & 39 s & 48 s\\
		\hline
 		1 & rrryyyrrryyy  & 6 s & 6 s\\
		\hline
 		2 & GGGrrrGGGrrr & 30 s & 39 s\\
		\hline
 		3 & yyyrrryyyrrr & 6 s & 6 s\\
 		\hline
	\end{tabular}
	\label{tlParam}
\end{table}

Moreover, we assume that traffic lights are time-gap actuated, i.e. a vehicle can activate it only if the previous actuation was at most $minGap$ seconds ago (3 seconds for our simulations) and the maximum phase duration is not exceeded.

Furthermore, the algorithm requires the knowledge of the travel time from the actuator to the corresponding intersection $j$ at speed limit: $T^{j}_{a-i} = \lceil \frac{D^{j}_{k}}{SL_{k}} \rceil$, where $D^{j}_{k}$ is the distance from the actuator on the incoming link $k$ to the intersection $j$ and $SL_{k}$ is the link's speed limit. Since the duration of the phase must be at least $minDuration^{j}$, the actuation must be enabled only after the time passes a specific threshold $T^{j}_{th} = minDuration^{j} - T^{j}_{a-i}$. The first vehicle must arrive within $minGap$ after the threshold to prolong the phase. If such a vehicle exists, the next car has $minGap$ seconds to trigger the TL again. The process terminates when either no such vehicle is found or the $maxDuration^{j}$ is reached.

\subsection{Speed Advisory System Modification}

Since we augment our prediction algorithm with a simplified version of SAS proposed in \cite{1}, we briefly summarize the main points of this system. The optimal (in terms of fuel consumption) speed trajectory consists of bang-singular-bang segments: (1) accelerate with maximal acceleration / decelerate with engine off - (2) keep the constant speed - (3) decelerate with engine off / accelerate with maximal acceleration. The singular segment is present only at very low speeds, so most of the time the optimal trajectory is bang-bang shaped. This is both hard to implement in real life and uncomfortable for drivers, so \cite{1} suggests a near-optimal speed trajectory: bang-singular ((1) accelerate with maximal acceleration / decelerate with engine off or with minimal deceleration - (2) keep the constant speed).

We further assume the acceleration and deceleration to be constant. According to the original dynamics, for speeds under $30 \sfrac{m}{s}$ the engine-off deceleration ranges from $0.1460 \sfrac{m}{s^2}$ to $0.1480 \sfrac{m}{s^2}$. The difference is insignificantly small and, therefore, rounding the deceleration up to $0.15 \sfrac{m}{s^2}$ would not make any noticeable impact on the system compared to other uncertainties and assumptions.
The acceleration also follows an almost linear pattern, so it was decided to set it to a constant value $a_{max}$ ($2.5 \sfrac{m}{s^2}$ in our model). In addition, it simplified the simulation implementation, since the model presented in SUMO uses constant acceleration.

\begin{figure}[ht]
  	\begin{subfigure}{0.49\linewidth}
    	\includegraphics[width=\linewidth]{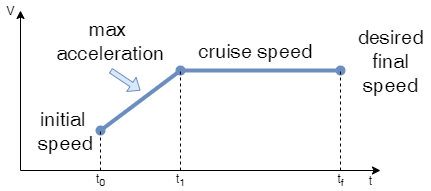}
     	\caption{Accelerate-then-cruise case.}
     	\label{fig3a}
  	\end{subfigure}
    \begin{subfigure}{0.49\linewidth}
    	\includegraphics[width=\linewidth]{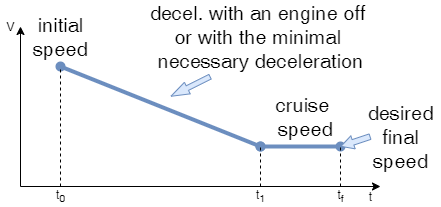}
    	\caption{Glide-then-cruise case.}
    	\label{fig3b}
  	\end{subfigure}
	\caption{Simplified near-optimal speed trajectories.}
	\label{subopt}
\end{figure}

The resulting near-optimal speed trajectory used in our simulation is one of the following (Fig. \ref{subopt}):
\begin{enumerate}
	\item \textbf{accelerate} with a constant maximal acceptable acceleration to a certain desired speed (not exceeding the speed limit) and \textbf{cruise},
	\item \textbf{decelerate} with an engine off ($\approx 0.15 \sfrac{m}{s^2}$) to a certain desired speed and \textbf{cruise},
	\item \textbf{apply} a necessary constant \textbf{braking} to meet boundary conditions.
\end{enumerate}

\subsection{Algorithm}

\subsubsection{Estimation of actuation time}

the moment a vehicle crosses the counter the algorithm estimates the time when this vehicle is going to reach the downstream actuator and maps it to the time within the current cycle.

\textbf{Step 1.} Compute the time required for the vehicle $i$ to travel from the counter to the actuator on the link $k$ (accelerate with $a_{max}$ to reach the speed limit and cruise):

\begin{equation*}
  	T^i_{travel} = \frac{SL_k - v_i}{a_{max}} + \frac{d_k - \frac{SL^2_k - v_i^2}{2a_{max}}}{SL_k},
\end{equation*}

\noindent where:

$v_i$ is the current speed of the vehicle $i$,

$d_k$ is the distance between the counter and the
actuator on the link $k$.

\textbf{Step 2.} Compute the ``counter" crossing time in the traffic light's frame of reference for the vehicle $i$: $T^i_{count} = T_{current} - T_{start}$, where $T_{start}$ is the time when the current phase started and $T_{current}$ is the current time of the day.
	
\textbf{Step 3.} Compute the estimated time within the phase when the vehicle $i$ is expected to arrive at the actuator: $T^j_{est} = T^i_{count} + T^i_{travel}$ and store it for one cycle for further computations.

\textbf{Remark 1:} This part of the algorithm is expected to be performed by the \emph{infrastructure}, more specifically, by the computer installed at the intersection.

\begin{figure}[ht]
  	\includegraphics[width=1.0\linewidth]{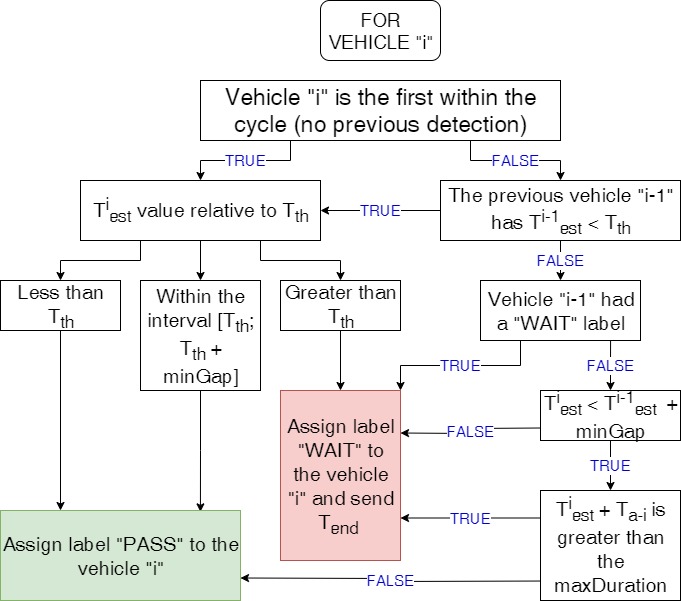}
  	\caption{Algorithm Diagram.}
  	\label{AlgDia}
\end{figure}

\subsubsection{``PASS" or ``WAIT" procedure} \label{pass}

after obtaining the estimated arrival time $T^i_{est}$ for the vehicle $i$, the algorithm proceeds to determine whether or not this vehicle will make it through the intersection within the current phase (Fig. \ref{AlgDia}). The algorithm iterates over the recorded $T_{est}$'s in search of the the first $T^a_{est}$ that lies within [$T^j_{th}; T^j_{th} + minGap$]. If the algorithm encounters $T^i_{est}$ before finding $T^a_{est}$ and $T^i_{est} < T^j_{th}$, the vehicle receives a ``PASS" label. If, however, $T^i_{est} > T^j_{th} + minGap$, the vehicle is ``WAIT" labeled. In case $T^a_{est}$ is reached first, the algorithm proceeds with the iteration checking if the $minGap$ is broken or $maxDuration^j$ is exceeded. If either of those occur before encountering $T^i_{est}$, the vehicle receives a ``WAIT" label. Otherwise, it receives a ``PASS" label. 
``PASS" label implies that the vehicle is expected to be able to pass the intersection within the current green phase. ``WAIT" label, in turn, suggests that the remaining time is insufficient for the vehicle to cross the intersection and advises it to wait for the next green phase.
 
Complete labeling of all vehicles allows the algorithm to derive an estimation for the current phase length:
\begin{equation*}
    T^j_{green} = max(T^*_{est} + T^j_{a-i}, minDuration^j),
\end{equation*}
where $T^*_{est}$ is the estimated arrival time for the last vehicle with ``PASS" label if any.

\textbf{Remark 2: }$T^j_{green}$ is not necessary for the near-optimal speed trajectory computation, but might be important for further development of the algorithm, giving vehicles on the ``secondary road" (non-actuated axis) an opportunity to construct their desired trajectories.

\textbf{Remark 3: }The proposed computations can be executed by either the \emph{infrastructure} or \emph{vehicles}. A detailed examination of these options can be found in Section \ref{distrib}.

\begin{figure}[ht]
  	\includegraphics[width=1.0\linewidth]{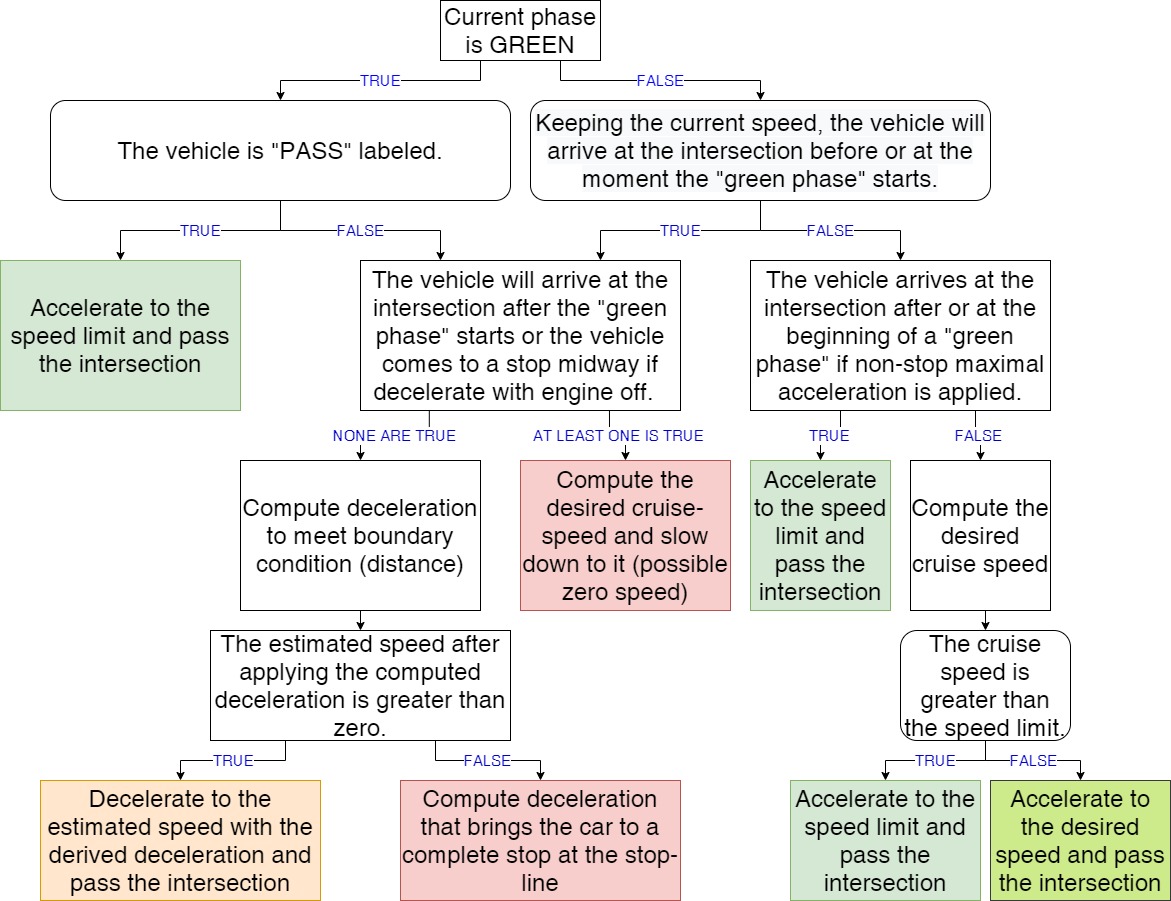}
  	\caption{Speed Advisory System Diagram.}
  	\label{SASDia}
\end{figure}

\subsubsection{Combining predictions and SAS}

at this stage, the near-optimal speed trajectory can be computed using the procedure illustrated in Fig. \ref{SASDia}. Vehicles labeled ``PASS" are advised to proceed as fast as possible to minimize their travel time. On the other hand, being labeled ``WAIT" is fundamentally equivalent to not being labeled at all (crossing the counter during a non-green phase). In both cases the residual time until the beginning of the next green phase is $T_{res} = cycLen_j - T^i_{count}$ and can be obtained directly from the infrastructure.

\textbf{Remark 4:} The near-optimal speed trajectory is derived independently by the Speed Advisory System installed at each participating \emph{vehicle}.

\section{Results}

The two main objectives of our algorithm are the correct prediction of the vehicles' passing capability and accurate phase residual time estimation for the Speed Advisory System. To test how well our approach meets these objectives, we simulate every vehicle with and without active SAS, and for each intersection on their path we compare the cycle numbers during which both versions of the vehicle crossed that intersection. Moreover, we evaluate the effectiveness of our algorithm by estimating fuel consumption in both cases and deriving the resulting gas savings. Vehicles without any driver-assistance system, which we refer to  as ``ordinary" vehicles, follow the Krauss car-following model.
	
\subsection{Simple case}
			
\subsubsection{Simulations}

to study various possible scenarios we conducted series of simulations featuring three different traffic demands: low demand ($\frac{1}{40} \frac{veh}{sec}$); medium demand $(\frac{1}{10} \frac{veh}{sec})$ and high demand $(\frac{1}{3} \frac{veh}{sec})$. Moreover, for every demand, different penetration rates of SAS-equipped vehicles were tested: 0\%, 20\%, 60\%, and 100\%. Examination of different combinations of penetration rates and demands not only allows us to compare the changes in fuel consumption, but also to study the impact of SAS-equipped vehicles on other traffic participants.

\subsubsection{Accuracy of ``PASS"-``WAIT" algorithm}

given the required traffic data, Speed Advisory System proposes the near-optimal speed trajectory (Fig. \ref{subopt}) that not only reduces fuel consumption but also results in minimal travel time. Therefore, if a vehicle has a chance to cross the intersection within the current phase, the algorithm must not advise it to stop and wait for the next green light. Failure to guide the vehicle through the intersection even though the vehicle is capable of crossing it is referred to as a ``mismatch". The simulation results demonstrating the number of mismatches are compiled in Tab. \ref{simMis}.

\begin{table}[ht]
    \caption{Cycle mismatches for various traffic demands.}
	\centering
	\begin{tabular}{ |P{2cm}||P{0.8cm}|P{0.8cm}|P{0.8cm}||P{2cm}|  }
 		\hline
 		\backslashbox[23mm]{Demand $(\frac{veh}{sec})$}{SAS \%} & 20\% & 60\% & 100\% & {} \\
 		\hline \hline
		\multirow{2}{*}{ Low ($\frac{1}{40}$)} & 58 & 176 & 285 &        \# Simulated Cars \\ \cline{2-5}
				& 0 & 0 & 0 &        \# Mismatch \\
		\hline
		\multirow{2}{*}{Medium ($\frac{1}{10}$)} & 176 & 565 & 960 &        \# Simulated Cars \\ \cline{2-5}
				& 1 & 1 & 3 &        \# Mismatch \\
 		\hline
		\multirow{2}{*}{High ($\frac{1}{3}$)} & 400 & 1251 & 2054 &        \# Simulated Cars \\ \cline{2-5}
				& 3 & 20 & 27 &        \# Mismatch \\
		\hline
	\end{tabular}
	\label{simMis}
\end{table}

The algorithm demonstrates 100\% prediction accuracy in free traffic, allowing all vehicles to pass the intersection within the earliest possible cycle.

For medium demand, we observe rare mismatches, which, however, are not caused by the miscalculation but rather by a model specification. Speed Advisory System treats yellow and red signals identically, disallowing intersection crossing during non-green phases. Ordinary vehicles, on the other hand, do not hesitate passing the intersection on yellow, creating a mismatch in our model.

The congested traffic brings more uncertainty to the prediction calculation making it less accurate. As a result, the number of errors increases; however, the accuracy remains significantly high: more than 98\%.

\begin{figure}[ht]
  	\begin{subfigure}{0.32\linewidth}
    	\includegraphics[width=\linewidth]{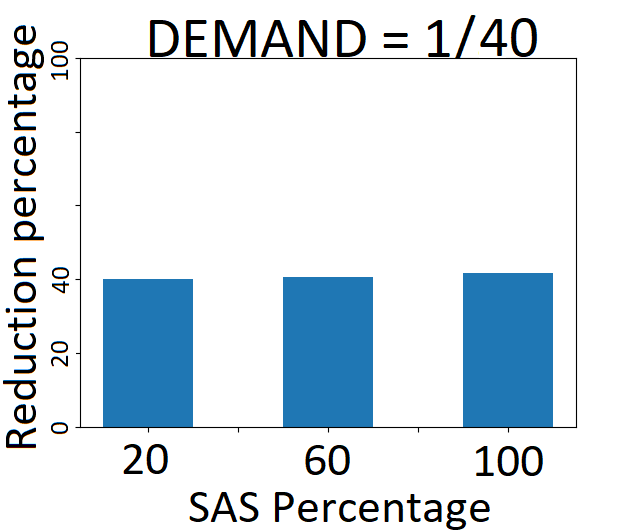}
     	\caption{Low demand.}
     	\label{simFuela}
  	\end{subfigure}
  	\hfill
  	\begin{subfigure}{0.32\linewidth}
    	\includegraphics[width=\linewidth]{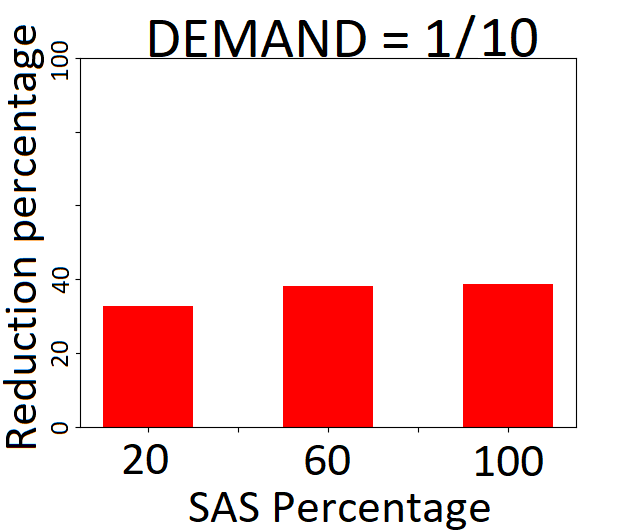}
    	\caption{Medium demand.}
    	\label{simFuelb}
  	\end{subfigure}
  	\hfill
  	\begin{subfigure}{0.32\linewidth}
	    \includegraphics[width=\linewidth]{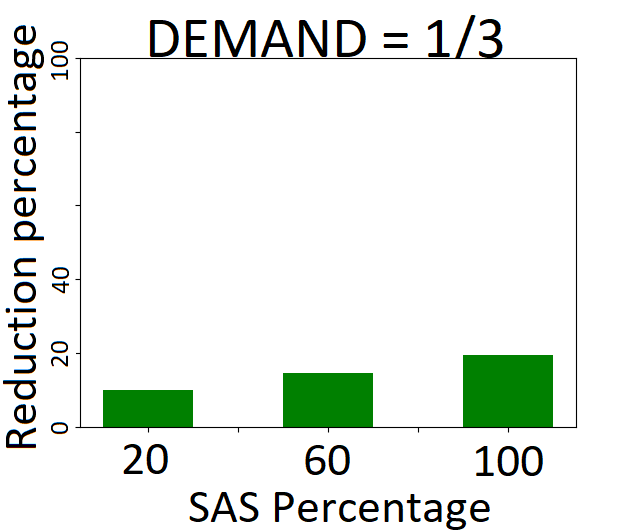}
    	\caption{High demand.}
    	\label{simFuelc}
  	\end{subfigure}
	\caption{Fuel consumption reduction for SAS-equipped vehicles in mixed traffic.}
	\label{simFuel}
\end{figure}

\subsubsection{Fuel Consumption}

the key benefit of the Speed Advisory System is fuel consumption reduction. By following near-optimal speed trajectories, vehicles manage to reduce idling at intersections and increase energy efficiency.
Introduction of prediction-based SAS demonstrates a significant improvement in fuel consumption for low and medium traffic demands: 35\% - 40\% (Fig. \ref{simFuela} - \ref{simFuelb}). According to Fig. \ref{simFuela}, the improvement is equivalent for all penetration rates of SAS-equipped vehicles in free traffic. The number of ordinary cars on the road is insufficient to affect speed profiles of controlled vehicles as long as the SAS penetration rate is above 20\% for the $\frac{1}{10} \frac{veh}{sec}$ demand scenario. In this case, however, the interference with the SAS-equipped vehicles becomes significant enough to slightly decrease fuel savings, but not enough to drastically deviate controlled vehicles from their desired speed patterns.

Furthermore, according to Fig. \ref{simFuelc}, congested traffic neutralizes most of the benefits of the Speed Advisory System. Queues forming due to excessive demand force vehicles to switch from the driver-assistance system to the car-following model, which negatively affects fuel consumption.

\begin{figure}[ht]
  	\begin{subfigure}{0.32\linewidth}
    	\includegraphics[width=\linewidth]{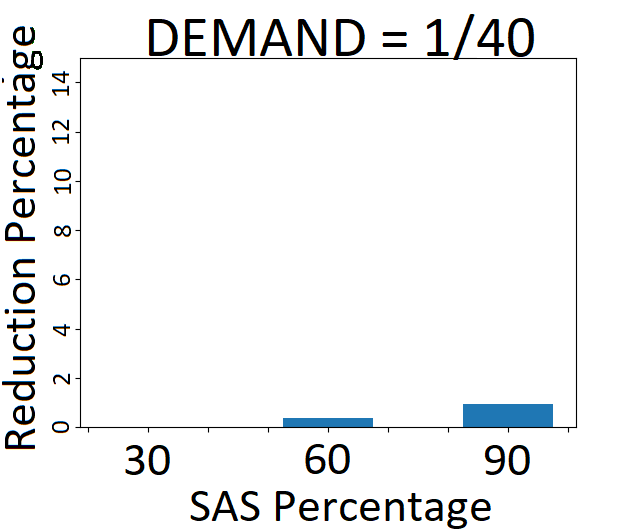}
     	\caption{Low demand.}
     	\label{simOrFuela}
  	\end{subfigure}
  	\hfill
  	\begin{subfigure}{0.32\linewidth}
    	\includegraphics[width=\linewidth]{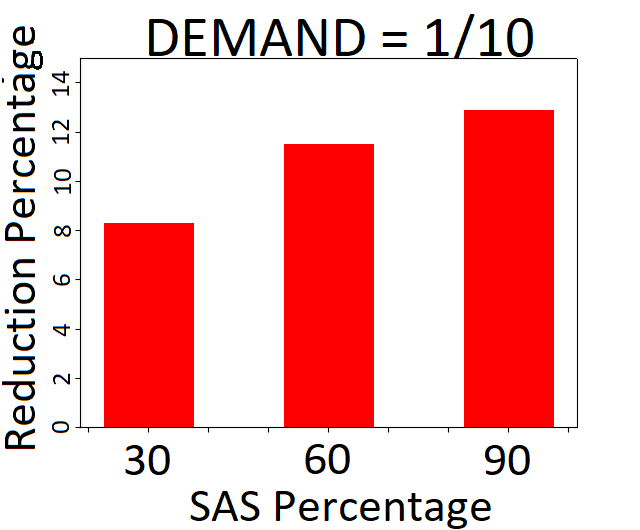}
    	\caption{Medium demand.}
    	\label{simOrFuelb}
  	\end{subfigure}
  	\hfill
  	\begin{subfigure}{0.32\linewidth}
	    \includegraphics[width=\linewidth]{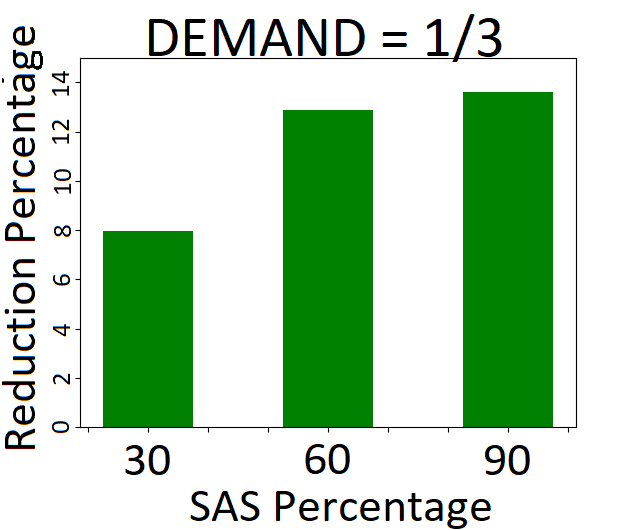}
    	\caption{High demand.}
    	\label{simOrFuelc}
  	\end{subfigure}
	\caption{Fuel consumption reduction for ordinary vehicles in mixed traffic.}
	\label{simOrFuel}
\end{figure}

In addition, we studied the impact the presence of SAS-equipped vehicles had on ordinary cars. We discovered that traffic participants with no driver-assistance system also manage to reduce fuel consumption, since they are forced to adjust their speed profiles to match the patterns proposed by surrounding controlled vehicles. In free traffic, the change is negligibly small: less than 1\% (Fig. \ref{simOrFuela}). However, in mildly and highly congested scenarios, the reduction is quite significant, ranging from 8.3\% to 12.5\% and from 8\% to 13.2\% respectively (Fig. \ref{simOrFuelb} - \ref{simOrFuelc}).

\subsubsection{Phase Utilization}

in addition to the primary objectives, several other performance measures were considered. First, we analyzed phase utilization, which can be effectively characterized by phase termination metric \cite{14}. There are four possible reasons for phase termination. An actuated phase can be \emph{omitted} when there is no actuation during the cycle; it can \emph{gap-out} when the TL was actuated at least once and then the actuation gap was broken; and it can \emph{max-out} when the phase duration reaches its maximum allowed length. Max-outs indicate that the phase is exceeding capacity, while gap-outs and omits indicate that there is capacity to spare. Our goal is to study the impact of the Speed Advisory System on the capacity utilization.

\begin{figure}[ht]
  	\begin{subfigure}{0.32\linewidth}
    	\includegraphics[width=\linewidth]{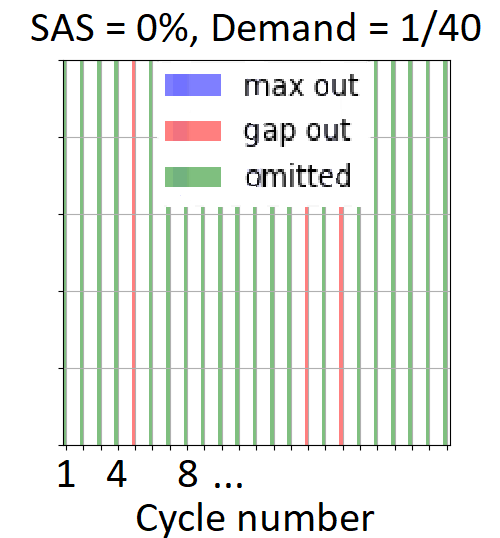}
  	\end{subfigure}
  	\hfill
  	\begin{subfigure}{0.32\linewidth}
    	\includegraphics[width=\linewidth]{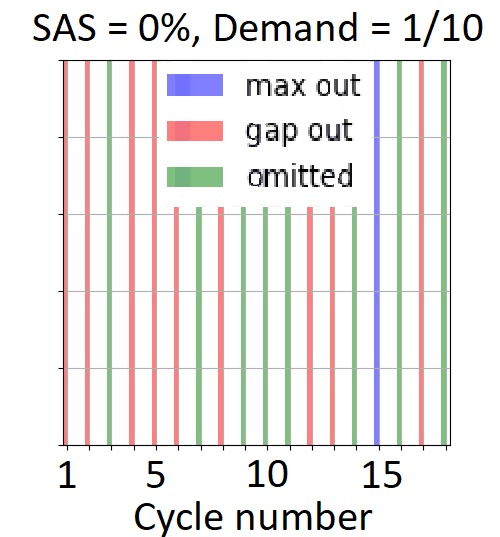}
  	\end{subfigure}
  	\hfill
  	\begin{subfigure}{0.32\linewidth}
	    \includegraphics[width=\linewidth]{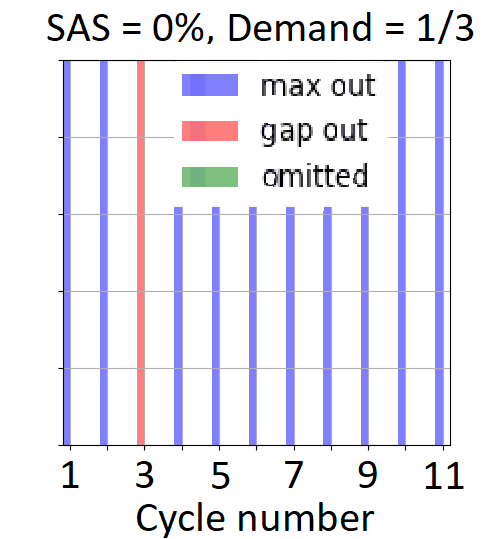}
  	\end{subfigure}
  	\hfill
	\begin{subfigure}{0.32\linewidth}
	    \includegraphics[width=\linewidth]{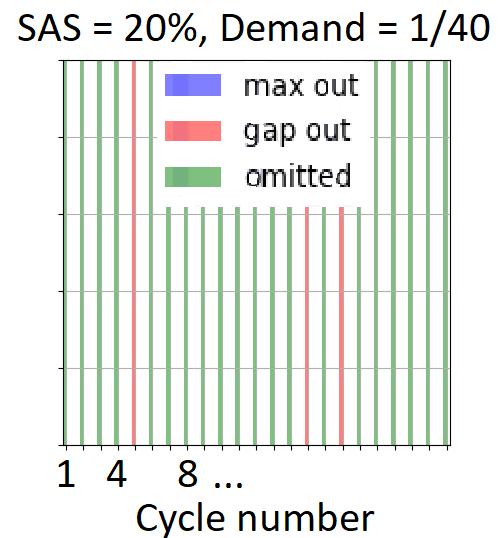}
  	\end{subfigure}
  	\hfill
	\begin{subfigure}{0.32\linewidth}
	    \includegraphics[width=\linewidth]{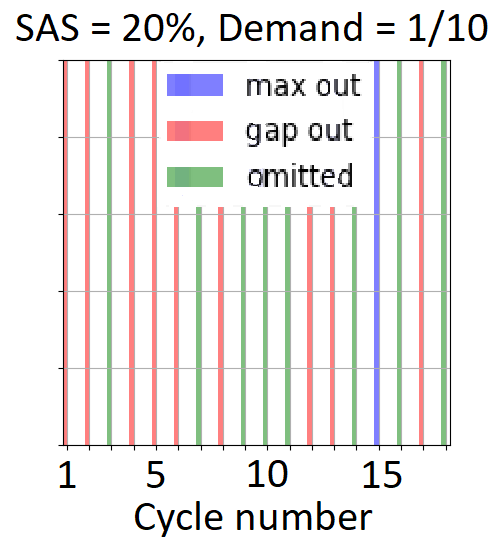}
  	\end{subfigure}
  	\hfill
	\begin{subfigure}{0.32\linewidth}
	    \includegraphics[width=\linewidth]{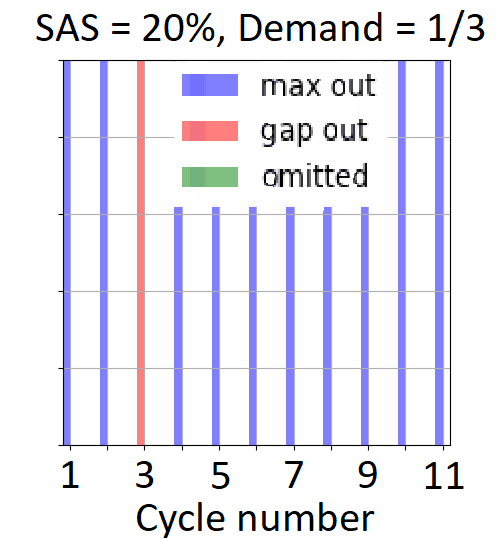}
  	\end{subfigure}
	\begin{subfigure}{0.32\linewidth}
	    \includegraphics[width=\linewidth]{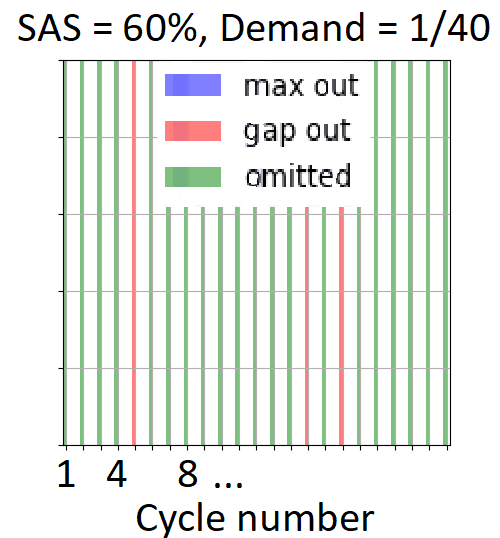}
  	\end{subfigure}
	\begin{subfigure}{0.32\linewidth}
	    \includegraphics[width=\linewidth]{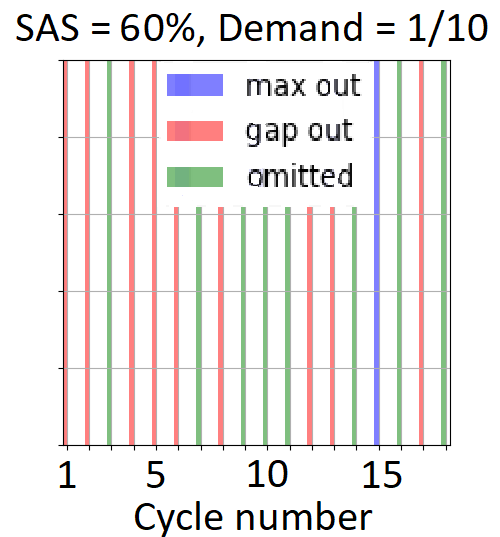}
  	\end{subfigure}
  	\hfill
	\begin{subfigure}{0.32\linewidth}
	    \includegraphics[width=\linewidth]{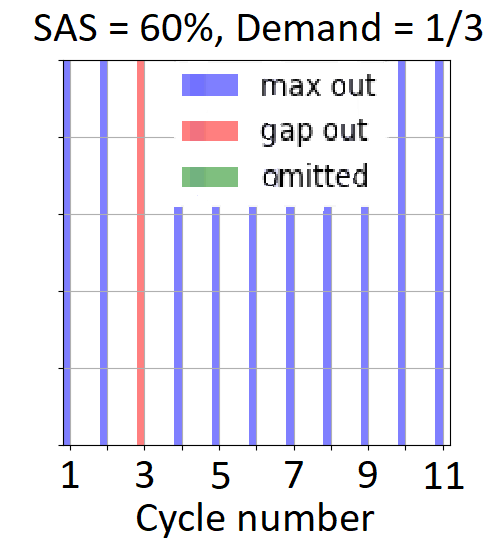}
  	\end{subfigure}
  	\hfill
	\begin{subfigure}{0.32\linewidth}
	    \includegraphics[width=\linewidth]{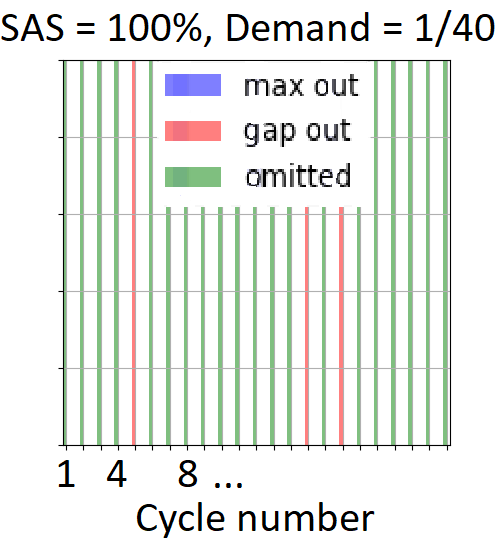}
		\caption{Low demand.}  			
	\end{subfigure}
	\hfill
	\begin{subfigure}{0.32\linewidth}
	    \includegraphics[width=\linewidth]{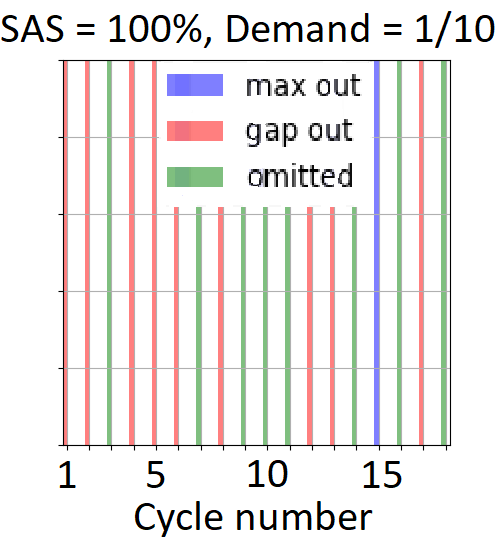}
		\caption{Medium demand.}
  	\end{subfigure}
  	\hfill
	\begin{subfigure}{0.32\linewidth}
	    \includegraphics[width=\linewidth]{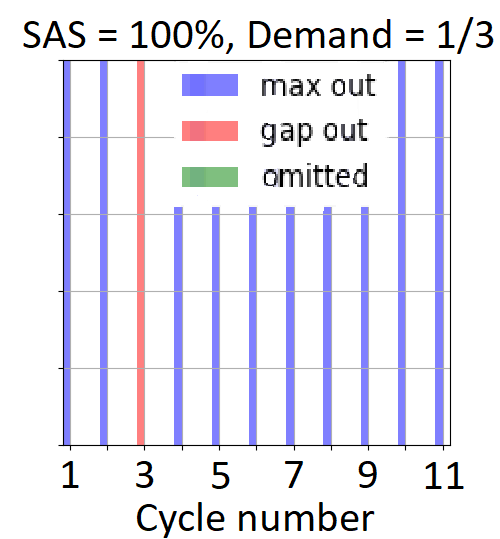}
    	\caption{High demand.}
  	\end{subfigure}
  	\hfill
	\caption{Phase termination reasons for various demands and SAS percentages.}
    \label{simPU}
\end{figure}

According to the simulation results (Fig. \ref{simPU}), driver-assistance system leaves the phase utilization unchanged for all levels of traffic demands. 

\subsubsection{Progression Quality}

another important performance measure is progression quality (PQ), which is strongly related to the queuing delay at an intersection due to arrival-departure patterns: high values of progression quality correspond to low delay.
We analyzed Percent-on-green (POG) values: $\frac{N_g}{N}$, where $N_g$ is the number of vehicles arriving during red and $N$ is the total number of vehicles arriving withing a cycle, to build an accurate representation of the PQ.


\begin{figure}[ht]
  	\begin{subfigure}{0.32\linewidth}
    	\includegraphics[width=\linewidth]{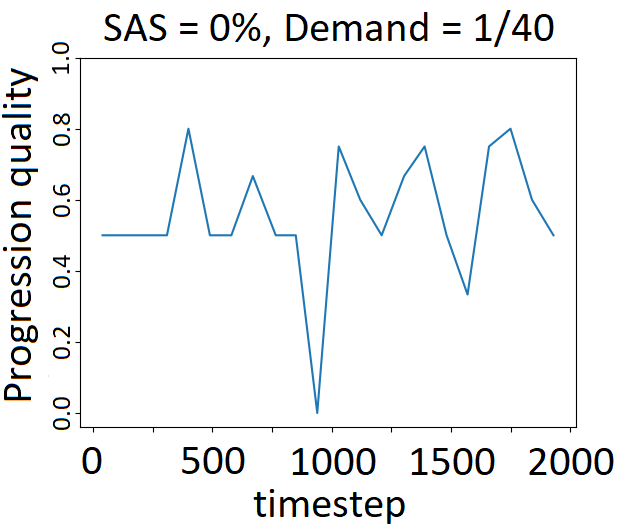}
  	\end{subfigure}
  	\hfill
  	\begin{subfigure}{0.32\linewidth}
    	\includegraphics[width=\linewidth]{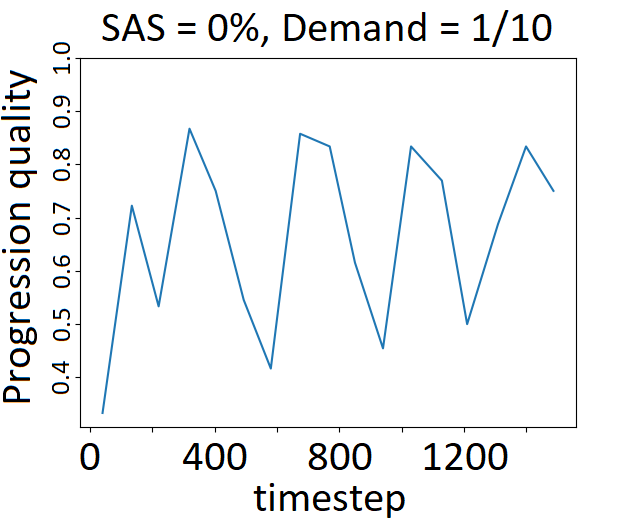}
  	\end{subfigure}
  	\hfill
  	\begin{subfigure}{0.32\linewidth}
	    \includegraphics[width=\linewidth]{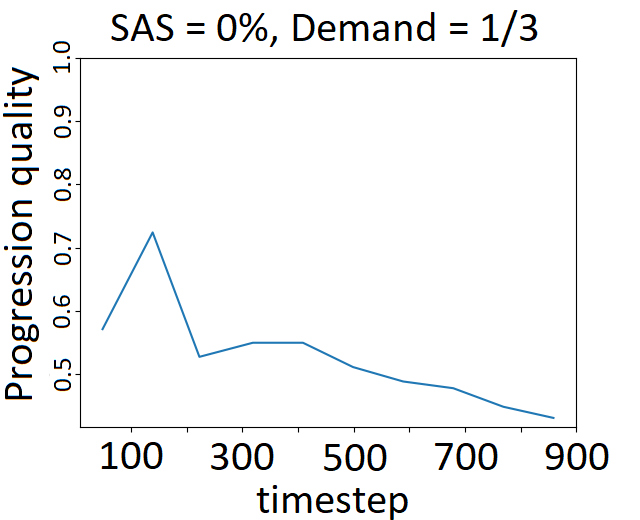}
  	\end{subfigure}
  	\hfill
	\begin{subfigure}{0.32\linewidth}
	    \includegraphics[width=\linewidth]{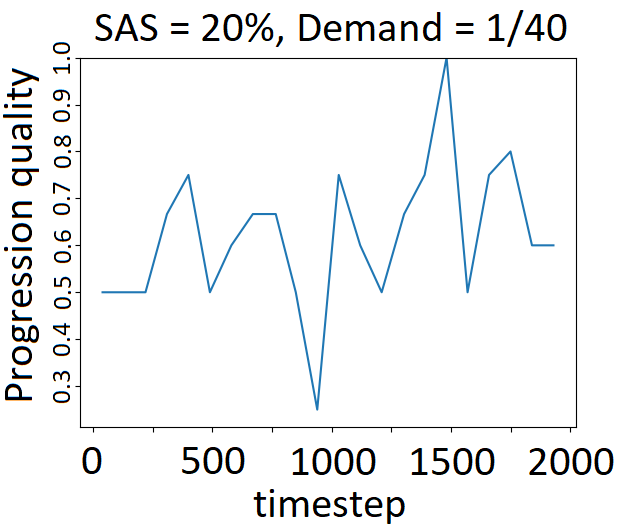}
  	\end{subfigure}
  	\hfill
	\begin{subfigure}{0.32\linewidth}
	    \includegraphics[width=\linewidth]{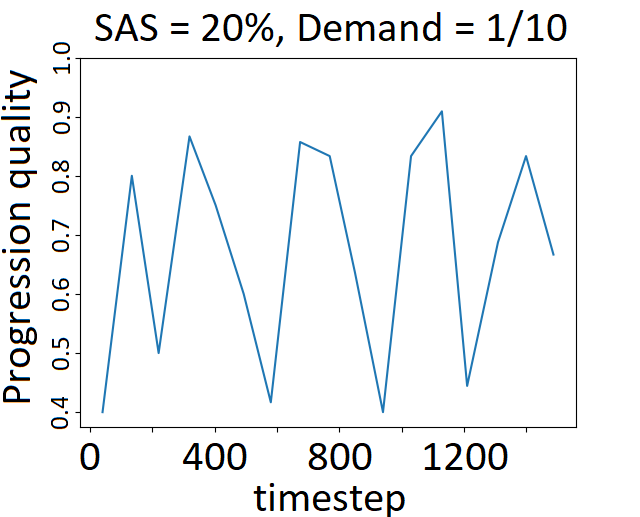}
  	\end{subfigure}
  	\hfill
	\begin{subfigure}{0.32\linewidth}
	    \includegraphics[width=\linewidth]{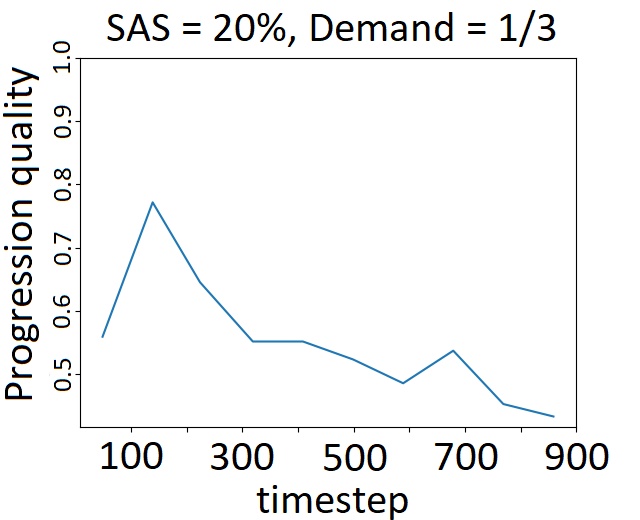}
  	\end{subfigure}
  	\hfill
	\begin{subfigure}{0.32\linewidth}
	    \includegraphics[width=\linewidth]{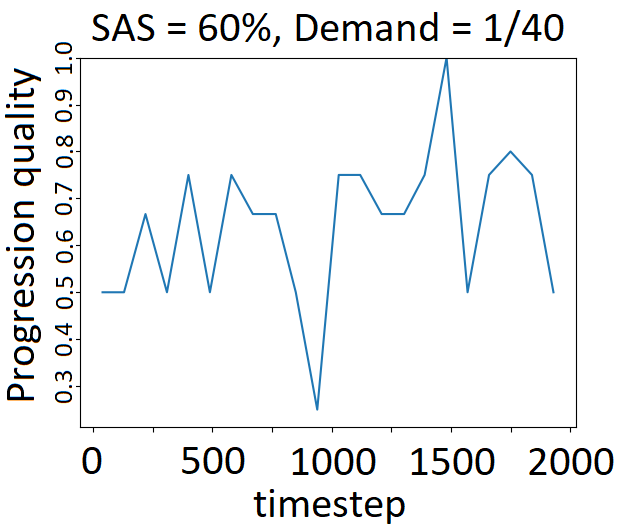}
  	\end{subfigure}
  	\hfill
	\begin{subfigure}{0.32\linewidth}
	    \includegraphics[width=\linewidth]{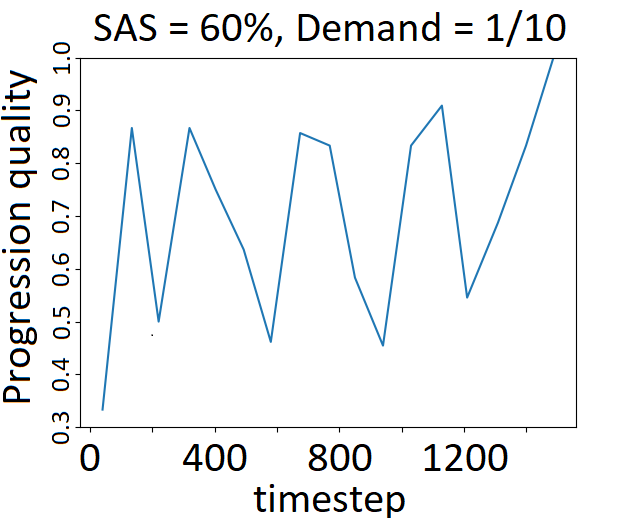}
  	\end{subfigure}
  	\hfill
	\begin{subfigure}{0.32\linewidth}
	    \includegraphics[width=\linewidth]{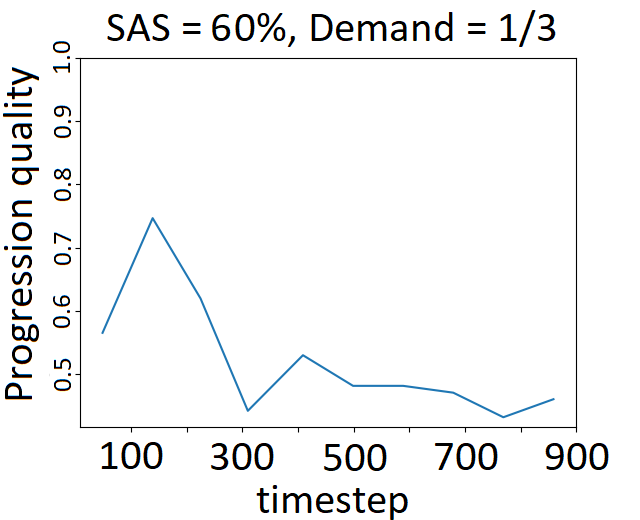}
  	\end{subfigure}
  	\hfill
	\begin{subfigure}{0.32\linewidth}
	    \includegraphics[width=\linewidth]{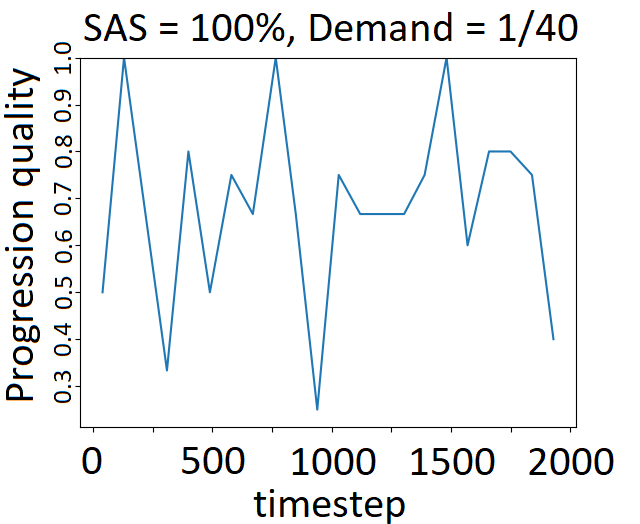}
		\caption{Low demand.}
		\label{simPCa}
	\end{subfigure}
	\hfill
	\begin{subfigure}{0.32\linewidth}
	    \includegraphics[width=\linewidth]{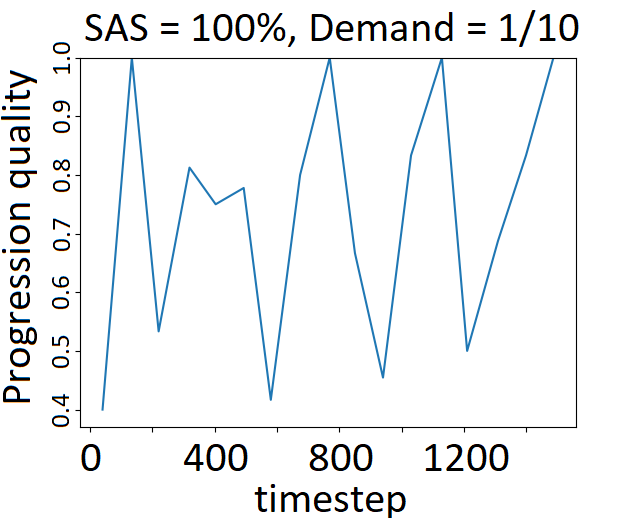}
		\caption{Medium demand.}
		\label{simPCb}
  	\end{subfigure}
  	\hfill
	\begin{subfigure}{0.32\linewidth}
	    \includegraphics[width=\linewidth]{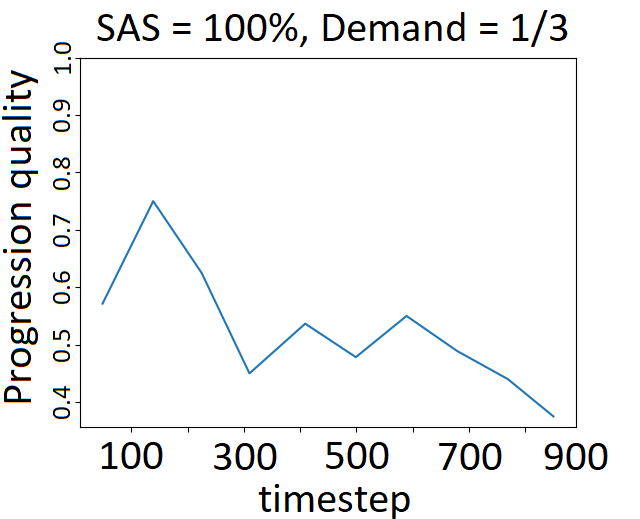}
    	\caption{High demand.}
    	\label{simPCc}
  	\end{subfigure}
  	\hfill
	\caption{Percent on green (POG) for progression quality measure.}
	\label{simPC}
\end{figure}

According to the low-demand simulation (Fig. \ref{simPCa}), we were able to increase the POG to 100\%, which corresponds to zero delay, for several cycles with 20\% and 100\% SAS penetration levels. Average and minimal POG values also benefited from the presence of connected vehicles.

In case of medium traffic demand (Fig. \ref{simPCb}), the algorithm demonstrated an improvement only for the 100\% SAS penetration rate scenario: three peaks of 100\% POG compared to zero peaks with no SAS introduced. For any other SAS penetration level, the progression quality is at least as good as the case with no SAS-equipped cars.

Finally, congested traffic (Fig. \ref{simPCc}), as expected, results in a relatively low progression quality, which can hardly be improved due to constant indissoluble or slowly dissoluble queues. In these conditions, Speed Advisory System is active for a short period of time before the vehicle switches to a car-following model and, therefore, has very limited opportunity to influence the vehicle's behavior.

\subsection{Montgomery County Network}

\subsubsection{Simulations}

due to the complex structure of the network it was possible to observe various traffic loads (free traffic, medium demand and congestion) in one setting. We tested three different SAS penetration levels: 0\%, 50\% and 100\%. In addition, three possible options for vehicles’ accelerations were implemented to analyze robustness of the algorithm: predetermined and fixed acceleration ($a = 2.5 \frac{m}{s^2}$), random but known acceleration and random unknown acceleration (for the last two scenarios $a$ is drawn from a uniform distribution in the interval $[2, 3.5]$). In the latter case the algorithm assumed that all vehicles had an average value of acceleration ($a = 2.75 \frac{m}{s^2}$).

\subsubsection{Accuracy}

the algorithm's prediction accuracy for 9 intersections is demonstrated in Table \ref{comAccur}. Green rows indicate free traffic, yellow rows correspond to moderate demand, and red rows are related to congestion. According to the data, the algorithm performs with an accuracy of at least 99\% in low demand for all possible acceleration scenarios. Having fewer vehicles on the road implies minimal interference, which, in turn, ensures precise calculations. 

\begin{table}[ht]
    \caption{``PASS" algorithm prediction accuracy}
	\centering
	\begin{tabular}{ |P{2cm}||P{1.1cm}|P{1.1cm}||P{2.6cm}|  }
 		\hline
 		\backslashbox[23mm]{Intersection}{SAS \%} & 50\% & 100\% & {} \\
 		\hline \hline				
		\rowcolor{Green!20}	& 100\% & 99.4\% &        Fixed known acc \\ \cline{2-4}
		\rowcolor{Green!20}	& 100\% & 99.4\% &											Random known acc \\ \cline{2-4}
		\rowcolor{Green!20}	\multirow{-3}{*}{ Intersection 1} & 100\% & 99.4\% & 	Random unknown acc \\		
		\hline
		\rowcolor{Green!20}	& 99.6\% & 99.8\% &											Fixed known acc \\ \cline{2-4}
		\rowcolor{Green!20}	& 99.6\% & 99.5\% &        									Random known acc \\ \cline{2-4}
		\rowcolor{Green!20}	\multirow{-3}{*}{ Intersection 2} & 99.3\% & 99.6\% &	Random unknown acc \\		
		\hline
		\rowcolor{Green!20}	& 98.7\% & 99.1\% &											Fixed known acc \\ \cline{2-4}
		\rowcolor{Green!20}	& 99\% & 99.2\% &        									Random known acc \\ \cline{2-4}
		\rowcolor{Green!20}	\multirow{-3}{*}{ Intersection 3}& 99\% & 99.2\% &		Random unknown acc \\		
		\hline
		\rowcolor{Yellow!20}	& 96.4\% & 98.7\% &									Fixed known acc \\ \cline{2-4}
		\rowcolor{Yellow!20}	& 98.3\% & 97.5\% &        							Random known acc \\ \cline{2-4}
		\rowcolor{Yellow!20}	\multirow{-3}{*}{ Intersection 4}& 98.1\% & 97.8\% &	Random unknown acc \\		
		\hline
		\rowcolor{Yellow!20}	& 98.5\% & 98.6\% &									Fixed known acc \\ \cline{2-4}
		\rowcolor{Yellow!20}	& 97.5\% & 97.5\% &									Random known acc \\ \cline{2-4}
		\rowcolor{Yellow!20}	\multirow{-3}{*}{ Intersection 5}& 97.7\% & 97.8\% &	Random unknown acc \\		
		\hline
		\rowcolor{Yellow!20}	& 89.3\% & 91.3\% &									Fixed known acc \\ \cline{2-4}
		\rowcolor{Yellow!20}	& 91.1\% & 91.3\% &        							Random known acc \\ \cline{2-4}
		\rowcolor{Yellow!20}	\multirow{-3}{*}{ Intersection 6}& 91.2\% & 91.3\% &	Random unknown acc \\		
		\hline
		\rowcolor{Red!20}	& 93.4\% & 92.5\% &											Fixed known acc \\ \cline{2-4}
		\rowcolor{Red!20}	& 93.2\% & 94.9\% &        								Random known acc \\ \cline{2-4}
		\rowcolor{Red!20}	\multirow{-3}{*}{ Intersection 7}& 94.6\% & 94.2\% &		Random unknown acc \\		
		\hline
		\rowcolor{Yellow!20}	& 97.9\% & 99.7\% &									Fixed known acc \\ \cline{2-4}
		\rowcolor{Yellow!20}	& 99.4\% & 99.7\% &        							Random known acc \\ \cline{2-4}
		\rowcolor{Yellow!20}	\multirow{-3}{*}{ Intersection 8}& 99.5\% & 99.7\% &	Random unknown acc \\		
		\hline
		\rowcolor{Yellow!20}	& 98\% & 98.1\% &										Fixed known acc \\ \cline{2-4}
		\rowcolor{Yellow!20}	& 99.3\% & 99.2\% &									Random known acc \\ \cline{2-4}
		\rowcolor{Yellow!20}	\multirow{-3}{*}{ Intersection 9}& 98.5\% & 99.3\% &	Random unknown acc \\		
		\hline
	\end{tabular}
	\label{comAccur}
\end{table}

Medium and high demands demonstrate at least 89\% accuracy in the worst case and $\approx 95\%$ on average. Vehicles interacting with each other are forced to slow down, accelerate or stop causing rare miscalculations. Moreover, intersection's geometry can have an impact on accuracy. Intersection 6 has relatively short incoming links, which makes the estimation of travel time difficult due to the discrete nature of the algorithm.

Furthermore, we compare the prediction errors of our algorithm (we will refer to it as ``Algorithm A") with those resulting from the statistically-based approach \cite{15}, which we refer to as ``Algorithm B". This approach is represented by an optimization problem, where phase duration estimation is addressed as a constraint of the form: $c_{p}^{i} \ge c_{r}^{i} + F^{-1}(\eta) $, where $c_{p}^{i}$  is the vehicle passing time in TL $i$ cycle clock, $c_{r}^{i}$ is minimal red-phase duration of the TL $i$, $F^{-1}$ is the inverse of the CDF function $F$ of random variable $\alpha$ representing stochastic time of delay, and $\eta$ is the required reliability level. The algorithm relies on the assumption that CDF is continuous and bijective. Moreover, the distribution function is non-parametric in general and may vary depending on arbitrary conditions. Although the paper presents impressive results in terms of fuel consumption (50\% - 57\%), the algorithm can be used only on the secondary road with no actuation capability (Effective Red implies that the perpendicular direction is the actuated one).

Although our algorithm does not provide the actual value of the green phase duration to the Speed Advisory System, we are able to estimate it based on the predicted arrival times, as discussed in section \ref{pass}. These estimated phase durations were used in comparison with the values provided by the Algorithm B for two confidence levels: $\eta = 0.8$ and $\eta = 0.1$.

 
		
\begin{figure}[ht]
  	\begin{subfigure}{0.49\linewidth}
    	\includegraphics[width=\linewidth]{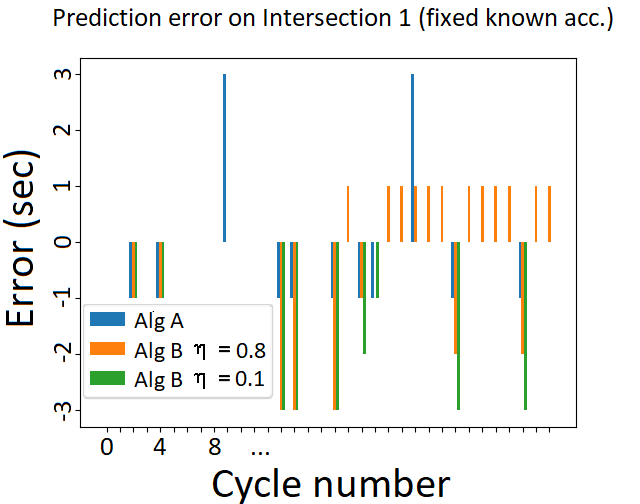}
  	\end{subfigure}
  	\hfill
  	\begin{subfigure}{0.49\linewidth}
	    \includegraphics[width=\linewidth]{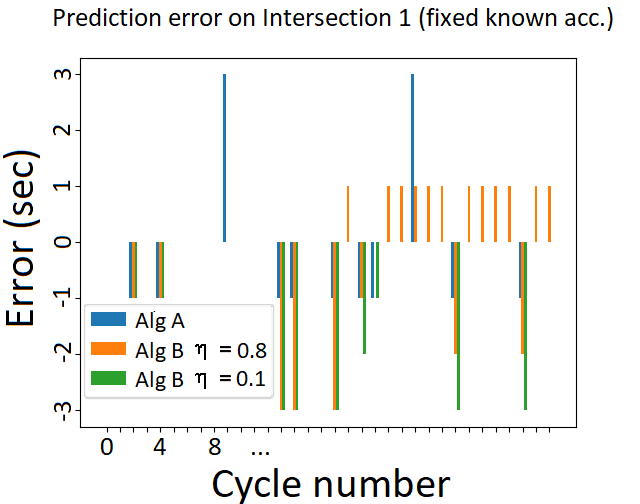}
	\end{subfigure}
	\hfill
  	\begin{subfigure}{0.49\linewidth}
    	\includegraphics[width=\linewidth]{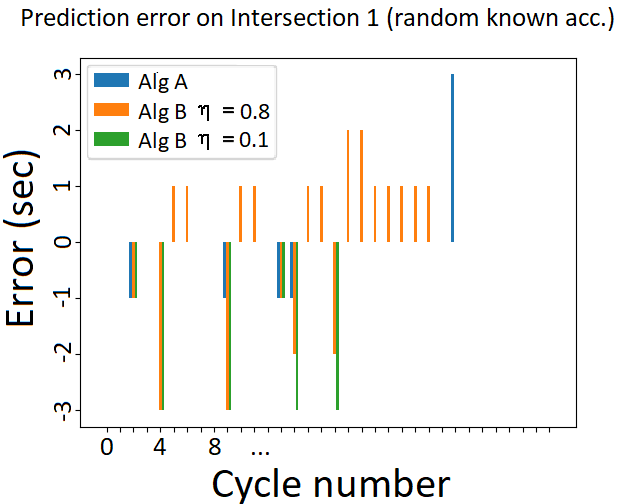}
  	\end{subfigure}
  	\hfill
  	\begin{subfigure}{0.49\linewidth}
	    \includegraphics[width=\linewidth]{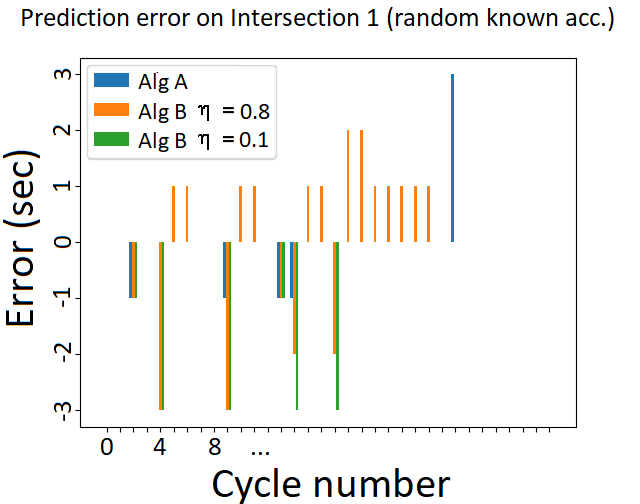}
	\end{subfigure}
	\hfill
  	\begin{subfigure}{0.49\linewidth}
	    \includegraphics[width=\linewidth]{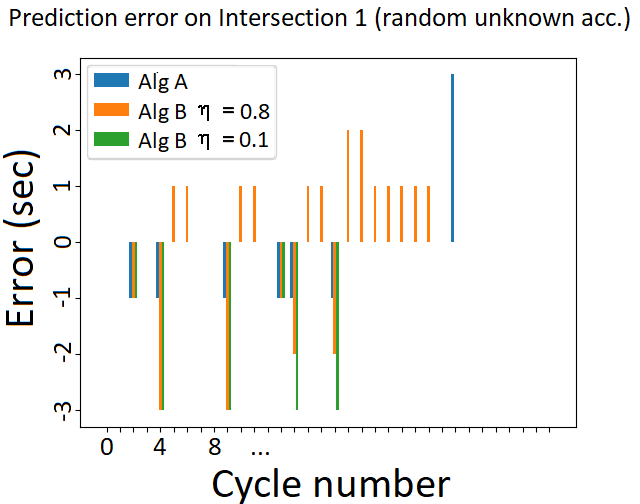}
	    \caption{100\% SAS penetration.}
  	\end{subfigure}
  	\hfill
	\begin{subfigure}{0.49\linewidth}
	    \includegraphics[width=\linewidth]{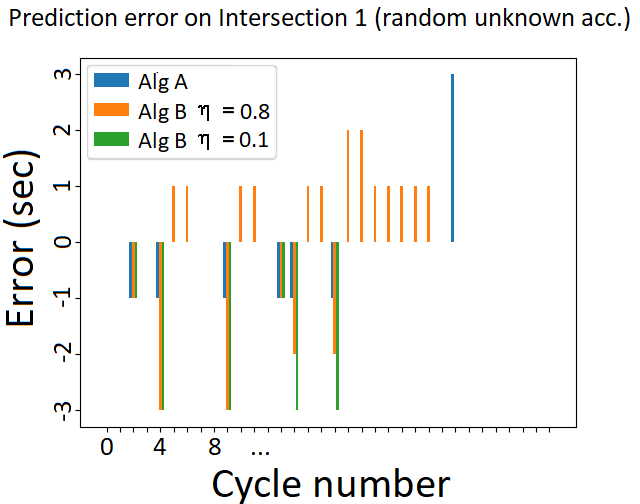}
  	\caption{50\% SAS penetrations.}
	\end{subfigure}
	\hfill
	\caption{Prediction errors in free traffic for Algorithm A and Algorithm B with $\eta=0.8$ and $\eta = 0.1$.}
    \label{comErr1}
\end{figure}

\begin{figure}[ht]
	\begin{subfigure}{0.49\linewidth}
	    \includegraphics[width=\linewidth]{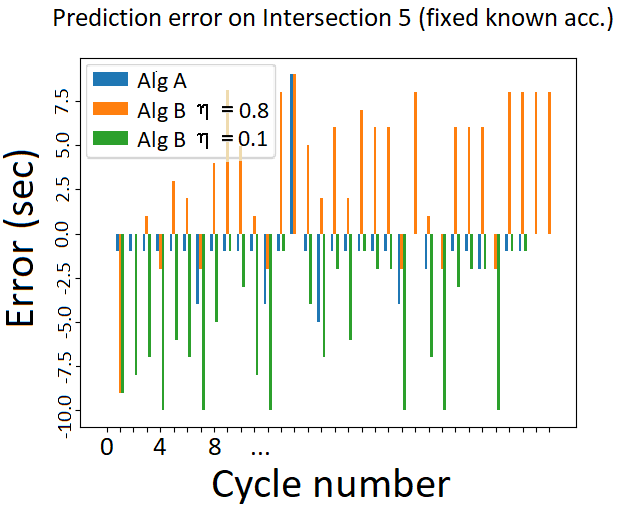}
  	\end{subfigure}
  	\hfill
  	\begin{subfigure}{0.49\linewidth}
	    \includegraphics[width=\linewidth]{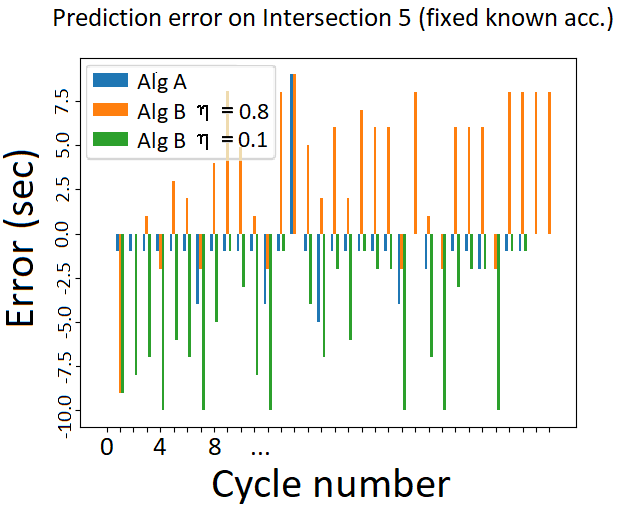}
	\end{subfigure}
	\hfill
	\begin{subfigure}{0.49\linewidth}
	    \includegraphics[width=\linewidth]{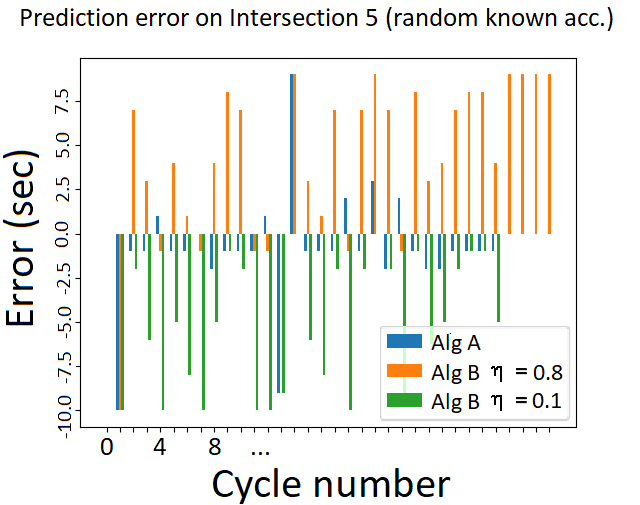}
  	\end{subfigure}
  	\hfill
  	\begin{subfigure}{0.49\linewidth}
	    \includegraphics[width=\linewidth]{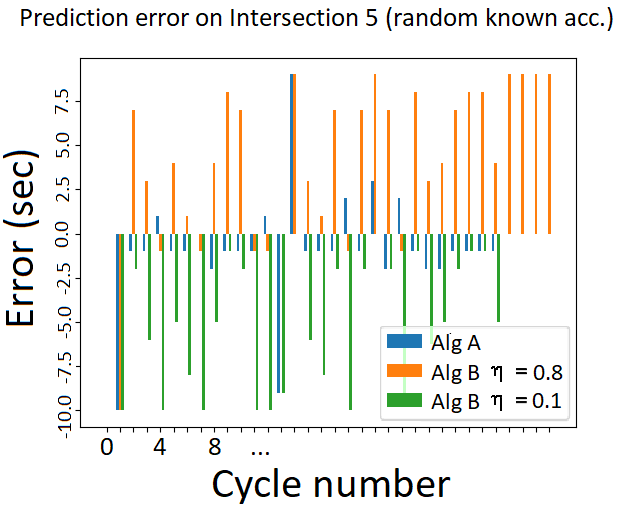}
  	\end{subfigure}
  	\hfill
	\begin{subfigure}{0.49\linewidth}
	    \includegraphics[width=\linewidth]{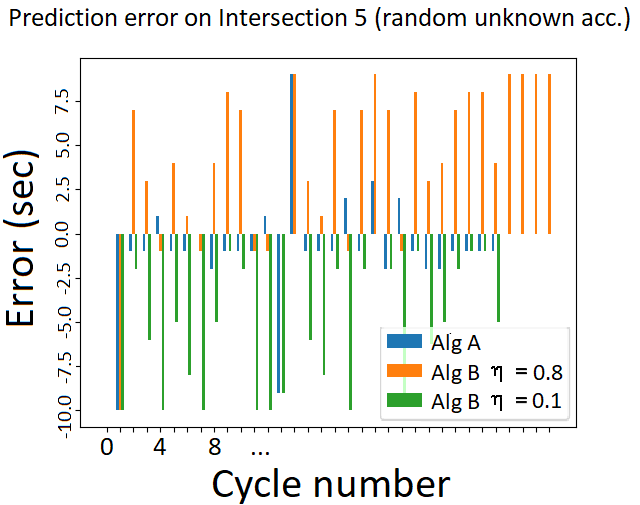}
	    \caption{100\% SAS penetration.}
  	\end{subfigure}
  	\hfill
	\begin{subfigure}{0.49\linewidth}
	    \includegraphics[width=\linewidth]{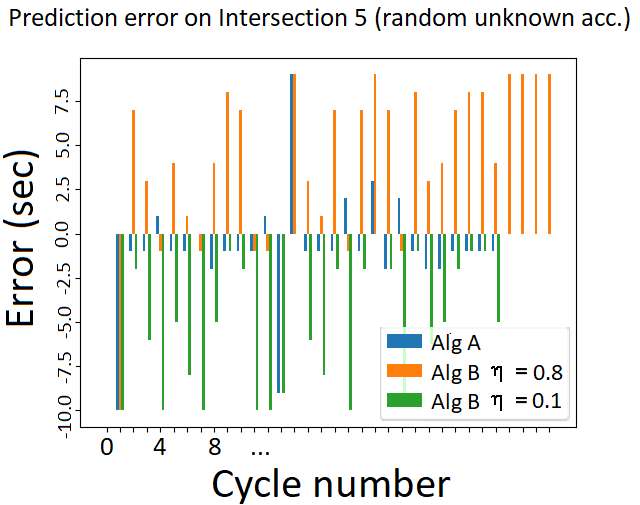}
    	\caption{50\% SAS penetration.}
  	\end{subfigure}
  	\hfill
	\caption{Prediction errors in medium demand for Algorithm A and Algorithm B with $\eta=0.8$ and $\eta=0.1$.}
    \label{comErr5}
\end{figure}

\begin{figure}[ht]
	\begin{subfigure}{0.49\linewidth}
	    \includegraphics[width=\linewidth]{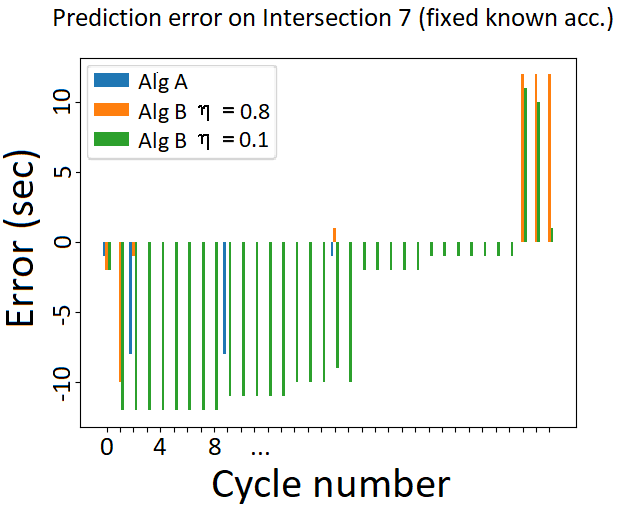}
	\end{subfigure}
	\hfill
	\begin{subfigure}{0.49\linewidth}
	    \includegraphics[width=\linewidth]{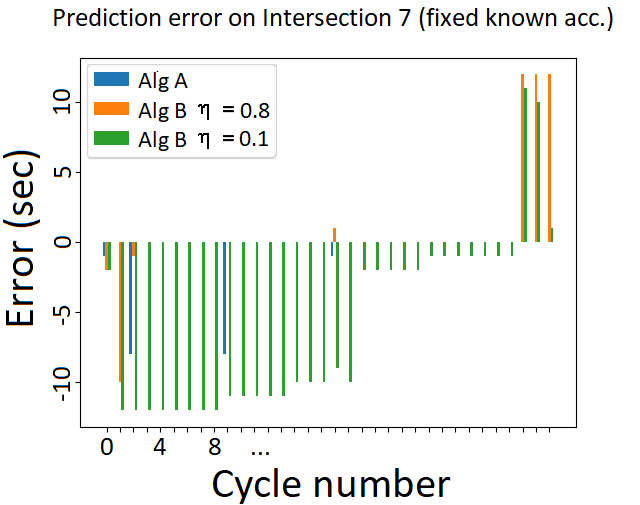}
	\end{subfigure}
	\hfill
	\begin{subfigure}{0.49\linewidth}
	    \includegraphics[width=\linewidth]{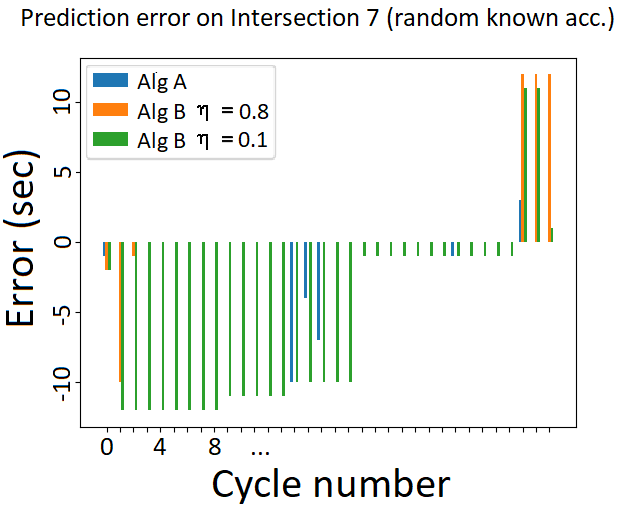}
  	\end{subfigure}
  	\hfill
  	\begin{subfigure}{0.49\linewidth}
	    \includegraphics[width=\linewidth]{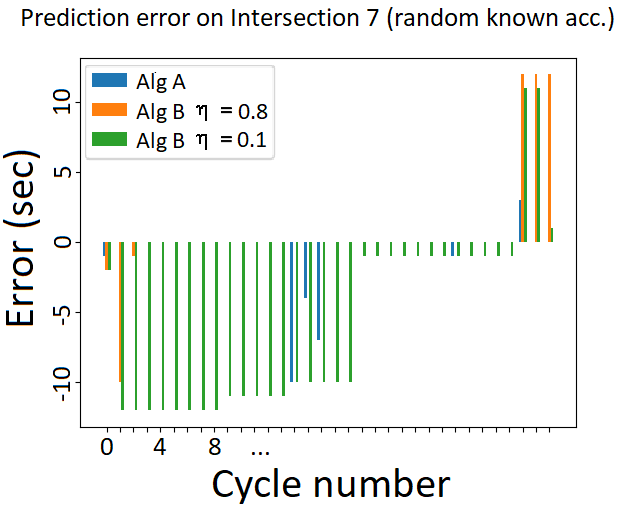}
  	\end{subfigure}
  	\hfill
	\begin{subfigure}{0.49\linewidth}
	    \includegraphics[width=\linewidth]{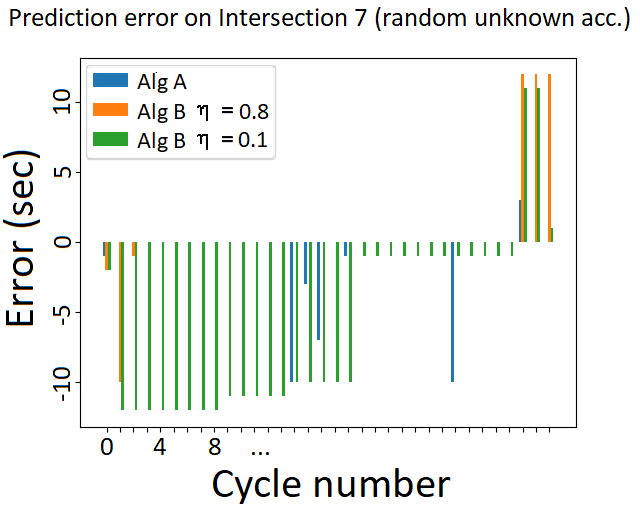}
	    \caption{100\% Penetration.}
  	\end{subfigure}
  	\hfill
	\begin{subfigure}{0.49\linewidth}
	    \includegraphics[width=\linewidth]{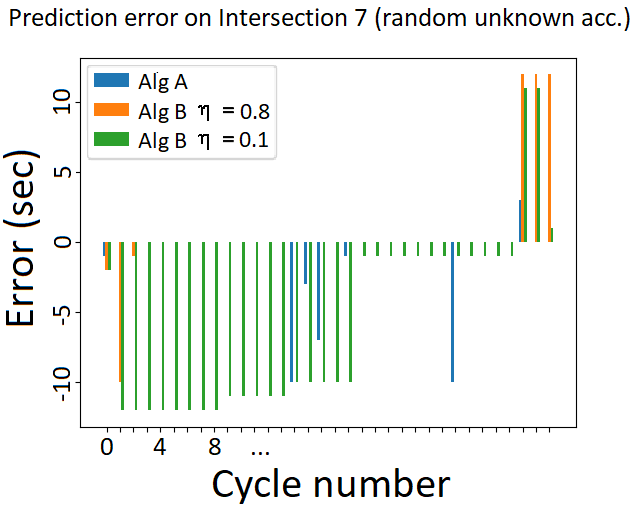}
    	\caption{50\% Penetration.}
  	\end{subfigure}
  	\hfill
	\caption{Prediction errors in high demand for Algorithm A and Algorithm B with $\eta=0.8$ and $\eta = 0.1$.}
    \label{comErr7}
\end{figure}

Fig. \ref{comErr1} - \ref{comErr7} demonstrate the difference (in seconds) between the predictions and the actual registered data for both algorithms. Positive and negative values of these errors correspond to overestimation and underestimation of the phase duration respectively. Each row of every figure corresponds to one of the three tested acceleration scenarios (top to bottom): fixed known accelerations, random known accelerations and random unknown accelerations.
We also assume that errors of less than 3 seconds are insignificant due to time discretization and vehicle dynamics simplification.

Fig. \ref{comErr1} contains data for the intersection 1 with low traffic demand. All three methods demonstrate high accuracy levels for every simulated scenario. Having fewer vehicles on the road implies low probability of triggering the traffic light actuation, which means that historical phase duration is almost always at minimal duration. Therefore, the CDF used in Algorithm B is almost constant and the prediction is relatively accurate. Although most of the errors can be viewed as insignificant, our method managed to precisely predict the phase duration more often than Algorithm B.

Medium traffic demand is represented by the intersection 5 (Fig. \ref{comErr5}). According to the histograms, the performance of our algorithm is much better than the performance of Algorithm B. Most of the errors produced by our method do not exceed 2 seconds compared to up to 10-second deviation for the statistically-based approach and occur less frequently. In the cases of random acceleration, greater errors appear, since Algorithm A has to rely on approximations of unavailable parameter values. However, these errors are relatively rare and have an insignificant impact on traffic flow. Regarding Algorithm B, setting the reliability level $\eta$ to 0.8 results in heavy overestimation for many cycles (up to 8-second errors). On the other hand, with $\eta = 0.1$ we obtain a serious underestimation of the phase length. In moderate traffic phase duration may vary from cycle to cycle with potentially high deviation. One additional data point does not change the CDF function significantly, and therefore, the phase length prediction for cycle $k$ will most likely be very similar to the prediction for cycle $k-1$ even if the actual phase durations differ dramatically.

Congested traffic data recorded at intersection 7 are compiled into the Fig. \ref{comErr7}. High demand resulted in  max-out termination (reaching the maximum allowed phase length) for most of the green phases during the simulation. 
The initial CDF for the Algorithm B corresponded to medium demand and required some time to receive enough data points to adjust and provide accurate predictions.  
Within the first several cycles, we observe poor performance for $\eta = 0.1$ with drastic underestimations up to 10 seconds. On the other hand, Algorithm A and Algorithm B with reliability level at 0.8 demonstrate impressive results with 100\% accuracy for most cycles. The terminal stage of the simulation corresponds to congestion dissolution and, as a result, to phase duration reduction. Algorithm B for both reliability levels cannot adapt to the sudden change in demand and continues to output maximum duration as the phase length prediction. Algorithm A, in comparison, manages to reflect the change in the traffic state and correctly estimate the new phase length. \emph{Statistically-based approach struggles to demonstrate sufficiently accurate results in the case of changing traffic demands, while the real-time algorithm that relies on the current traffic data is able to quickly adapt and accurately predict the phase length}.

\begin{figure}[ht]
	\begin{subfigure}{0.49\linewidth}
	    \includegraphics[width=\linewidth]{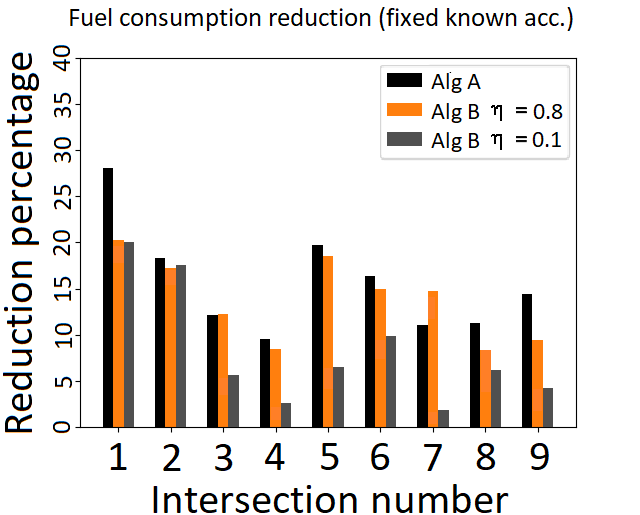}
  	\end{subfigure}
  	\hfill
  	\begin{subfigure}{0.49\linewidth}
	    \includegraphics[width=\linewidth]{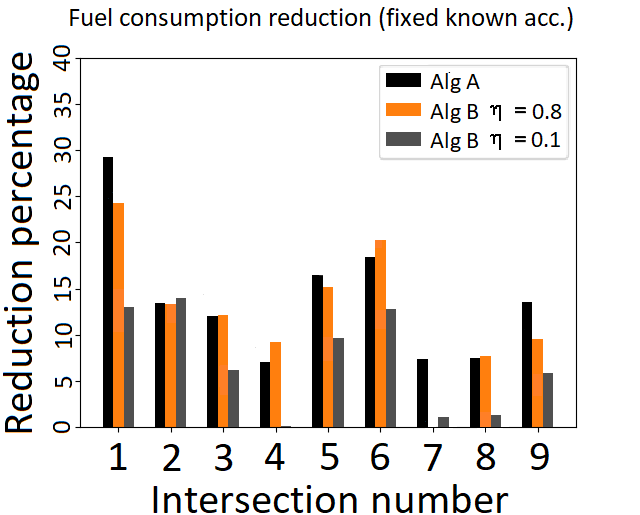}
	\end{subfigure}
	\hfill
	\begin{subfigure}{0.49\linewidth}
	    \includegraphics[width=\linewidth]{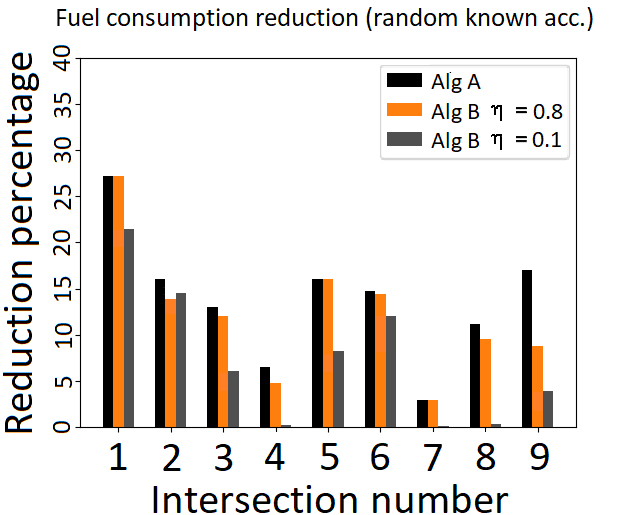}
  	\end{subfigure}
  	\hfill
  	\begin{subfigure}{0.49\linewidth}
	    \includegraphics[width=\linewidth]{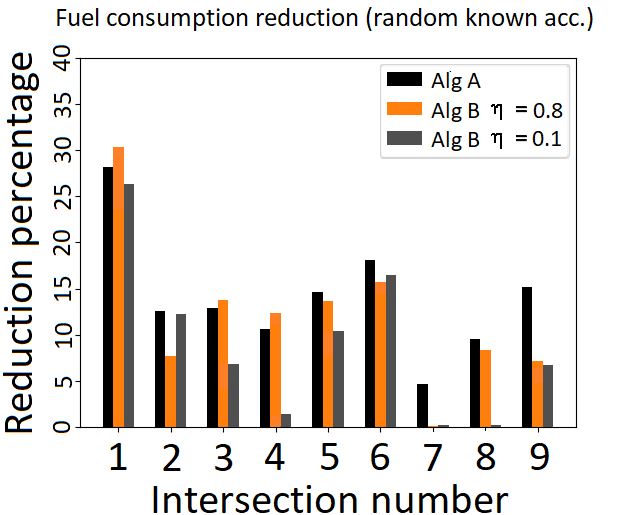}
  	\end{subfigure}
  	\hfill
	\begin{subfigure}{0.49\linewidth}
	   	\includegraphics[width=\linewidth]{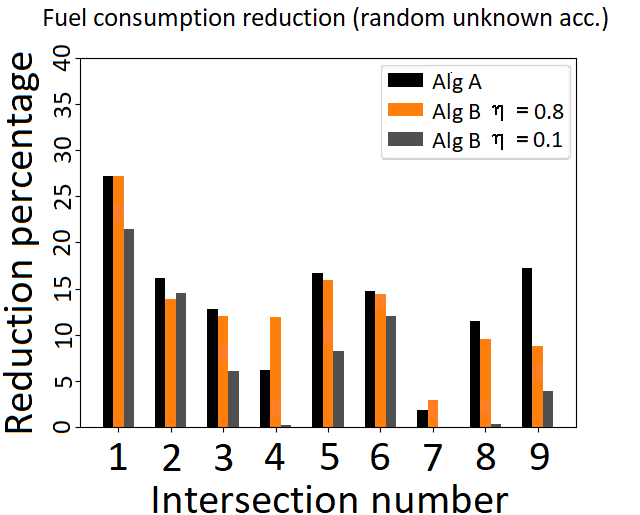}
	   	\caption{100\% Penetration.}
  	\end{subfigure}
  	\hfill
	\begin{subfigure}{0.49\linewidth}
	    \includegraphics[width=\linewidth]{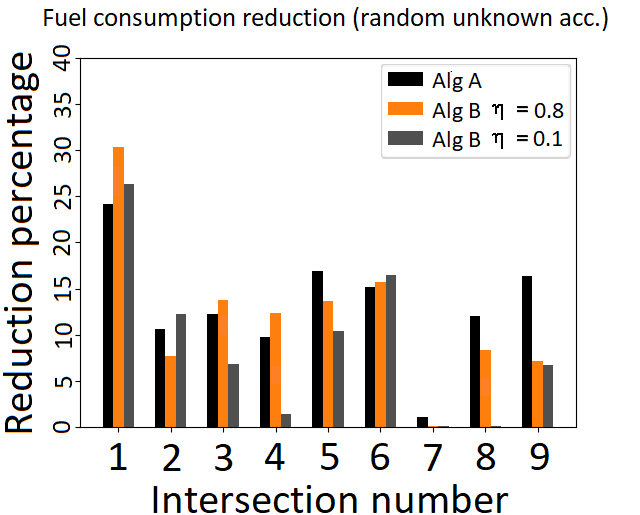}
    	\caption{50\% Penetration.}
  	\end{subfigure}
  	\hfill
	\caption{Fuel consumption reduction for Algorithm A and Algorithm B with $\eta=0.8$ and $\eta=0.1$ in mixed traffic.}
    \label{comFuel}
\end{figure}

\subsubsection{Fuel Consumption}

we also compared the impact on fuel consumption for both algorithms (Fig. \ref{comFuel}). According to the histograms, Algorithm A performs better on average than Algorithm B for both tested reliability levels. Setting $\eta$ to 0.1 results in much smaller fuel consumption reduction for all intersections and traffic demands. Moreover, choosing the reliability level to be 0.8 for Algorithm B results in a similar pattern as applying our algorithm. Switching between scenarios with different acceleration settings and SAS-vehicle penetration rates does not provide a significantly distinct outcome. In rare cases some intersections benefit more from Algorithm B; however the difference in savings is small.

Next step is to analyze our algorithm's performance independently of any other procedure. We managed to achieve up to 29\% fuel consumption reduction for free traffic set-ups. Medium and high demand scenarios also benefit from the presence of real-time prediction algorithm resulting in up to 18\% and 7\% fuel savings respectively. These results correlate with the simple case simulations. The Speed Advisory System is most effective in terms of fuel consumption in free and moderate traffic, because vehicles are less likely to be distracted from following the proposed near-optimal trajectories due to low interaction rates. Congestion, on the other hand, forces traffic participants to switch from SAS to car-following model and lose most of the impact driver-assistance could have on energy savings.

\begin{figure}[ht]
	\begin{subfigure}{0.32\linewidth}
	    \includegraphics[width=\linewidth]{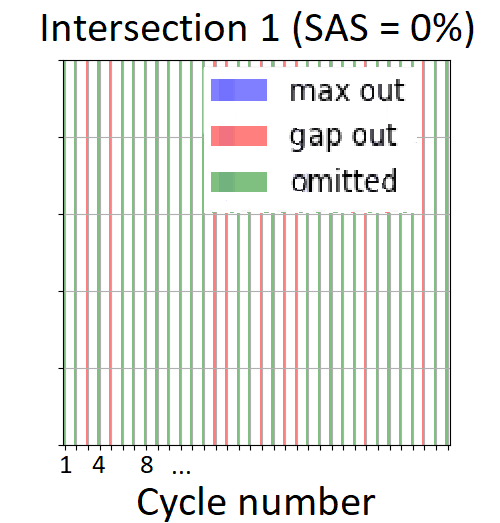}
  	\end{subfigure}
  	\hfill
	\begin{subfigure}{0.32\linewidth}
	    \includegraphics[width=\linewidth]{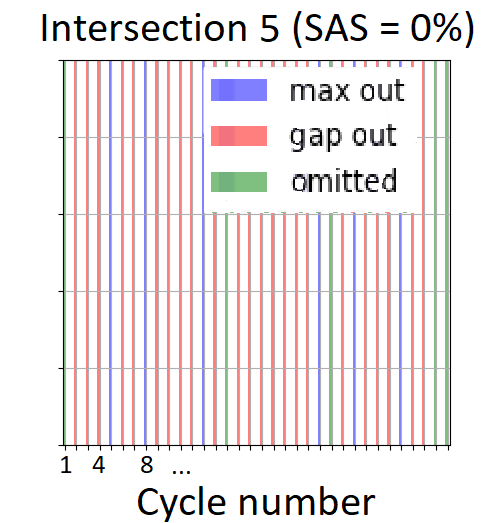}
  	\end{subfigure}
  	\hfill
	\begin{subfigure}{0.32\linewidth}
	    \includegraphics[width=\linewidth]{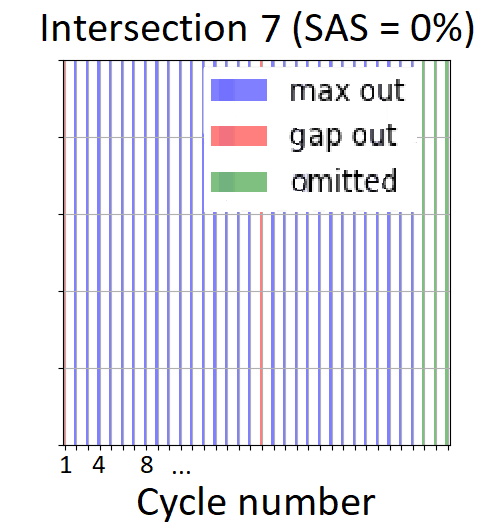}
  	\end{subfigure}
  	\hfill
	\begin{subfigure}{0.32\linewidth}
	    \includegraphics[width=\linewidth]{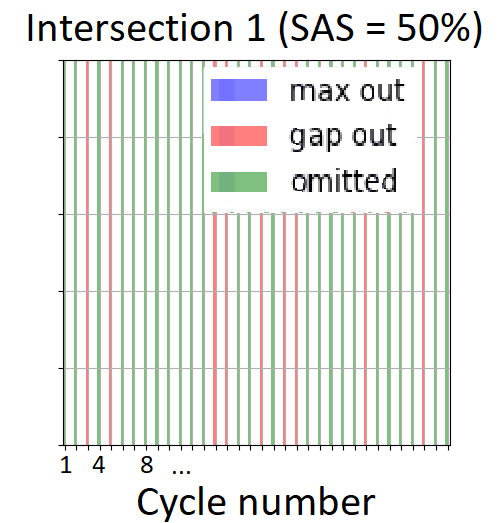}
  	\end{subfigure}
  	\hfill
	\begin{subfigure}{0.32\linewidth}
	    \includegraphics[width=\linewidth]{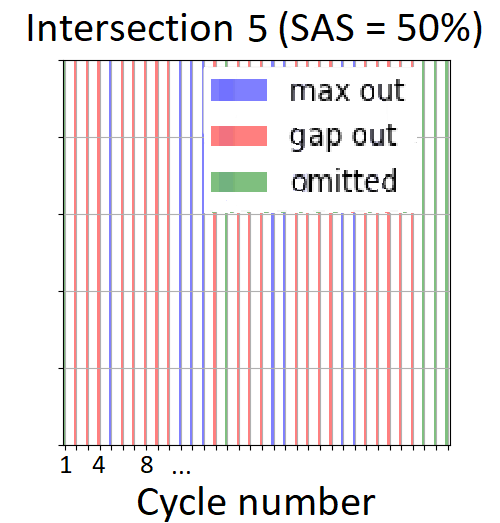}
  	\end{subfigure}
  	\hfill
	\begin{subfigure}{0.32\linewidth}
	    \includegraphics[width=\linewidth]{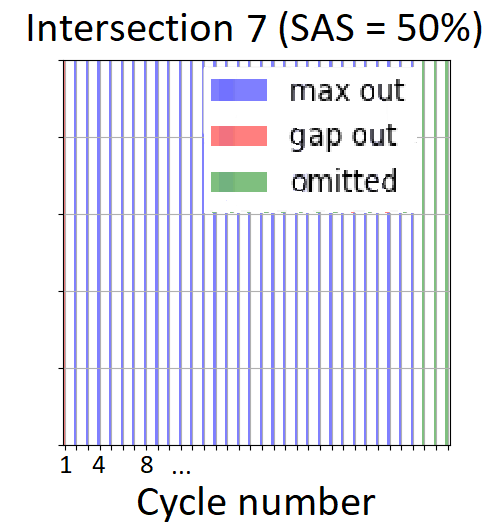}
  	\end{subfigure}
  	\hfill
	\begin{subfigure}{0.32\linewidth}
	    \includegraphics[width=\linewidth]{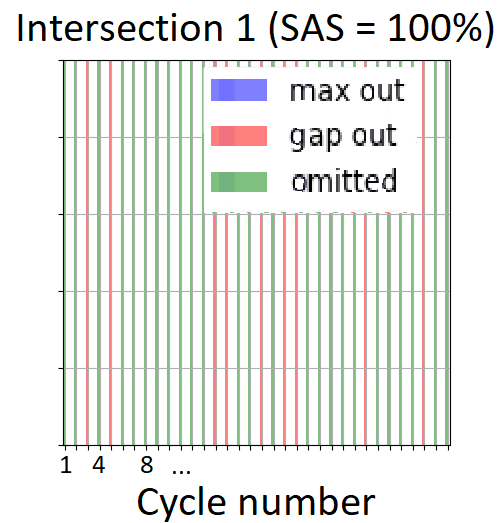}
		\caption{Low demand}
	\end{subfigure}
	\hfill
	\begin{subfigure}{0.32\linewidth}
	    \includegraphics[width=\linewidth]{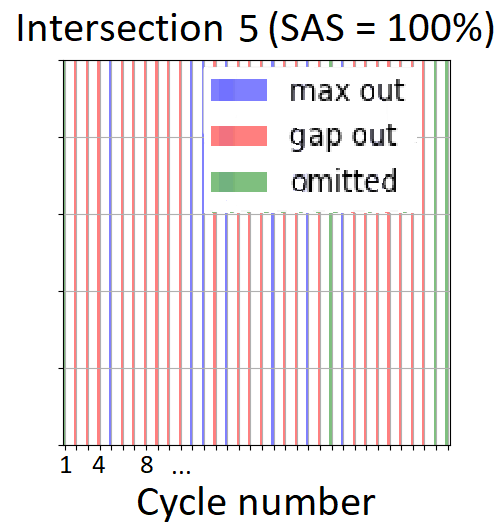}
		\caption{Medium demand}
		\end{subfigure}
  	\hfill
	\begin{subfigure}{0.32\linewidth}
	    \includegraphics[width=\linewidth]{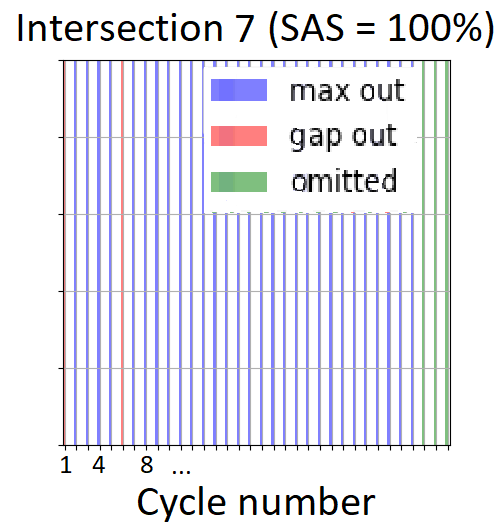}
    	\caption{High demand}
  	\end{subfigure}
  	\hfill
	\caption{Phase utilization for different SAS-equipped vehicle penetration levels.}
    \label{comPU}
\end{figure}

\subsubsection{Phase utilization}

we present the results for only known fixed acceleration scenario, since the other setups (known random and unknown random) produce similar outcomes. The data from intersections 1, 5 and 7 are compiled into Fig. \ref{comPU}. The first, second and third rows corresponds to 0\%, 50\% and 100\% SAS-equipped vehicle penetration rates respectively. 

Similarly to the simple case, phase utilization is barely affected by the introduction of the Speed Advisory System. Therefore, we can conclude that even if the difference is insignificant, the Speed Advisory System does not harm the traffic state of the network.

\begin{figure}[ht]
	\begin{subfigure}{0.32\linewidth}
	    \includegraphics[width=\linewidth]{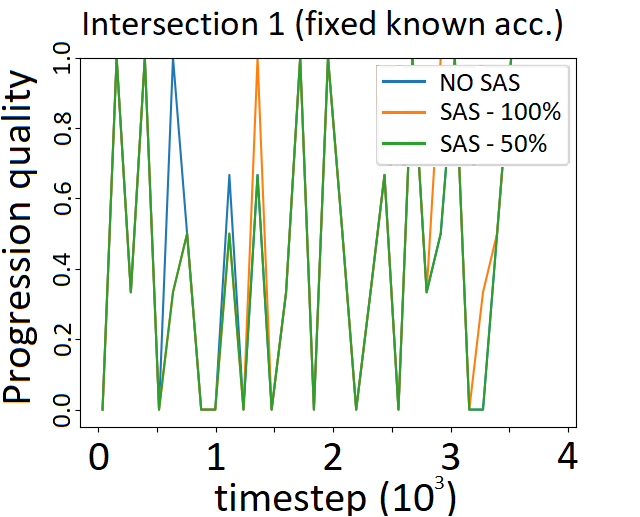}
  	\end{subfigure}
  	\hfill
	\begin{subfigure}{0.32\linewidth}
	    \includegraphics[width=\linewidth]{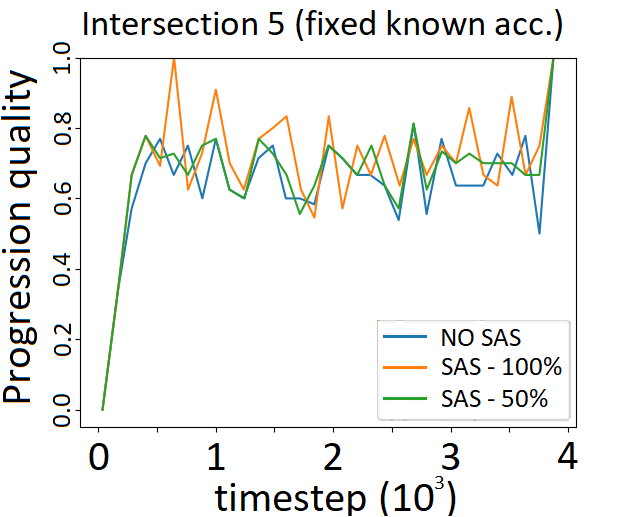}
  	\end{subfigure}
  	\hfill
	\begin{subfigure}{0.32\linewidth}
	    \includegraphics[width=\linewidth]{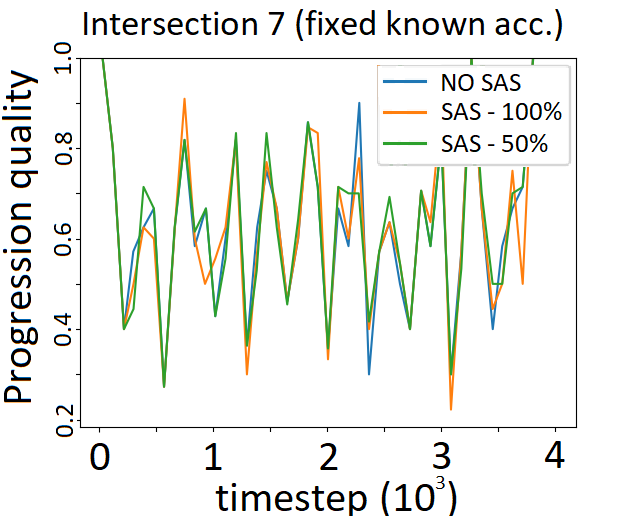}
  	\end{subfigure}
  	\hfill
	\begin{subfigure}{0.32\linewidth}
	    \includegraphics[width=\linewidth]{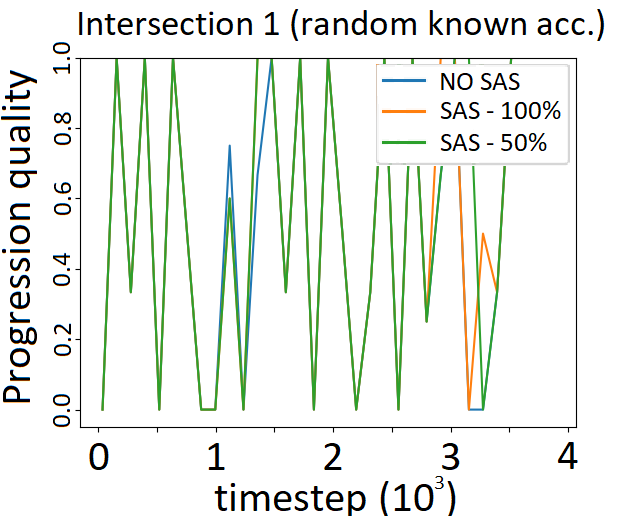}
  	\end{subfigure}
  	\hfill
	\begin{subfigure}{0.32\linewidth}
	    \includegraphics[width=\linewidth]{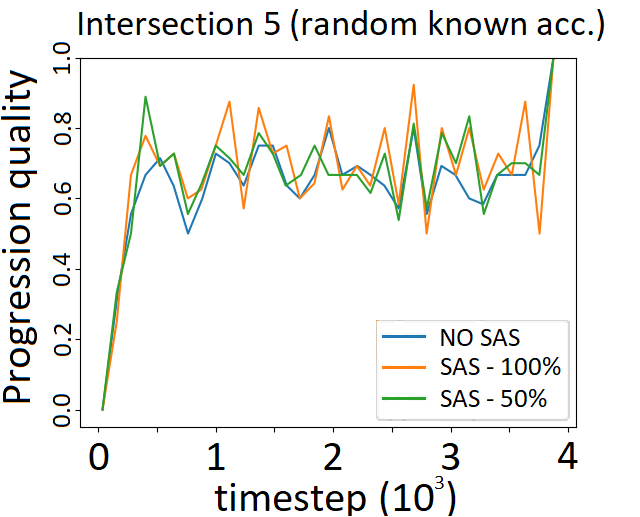}
  	\end{subfigure}
  	\hfill
	\begin{subfigure}{0.32\linewidth}
	    \includegraphics[width=\linewidth]{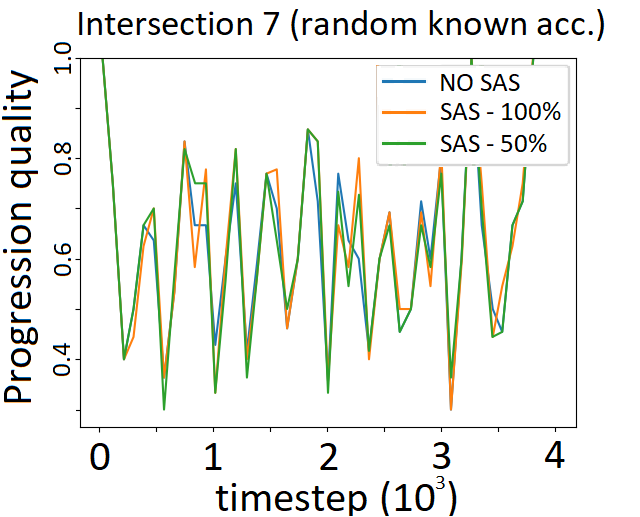}
  	\end{subfigure}
  	\hfill
	\begin{subfigure}{0.32\linewidth}
	    \includegraphics[width=\linewidth]{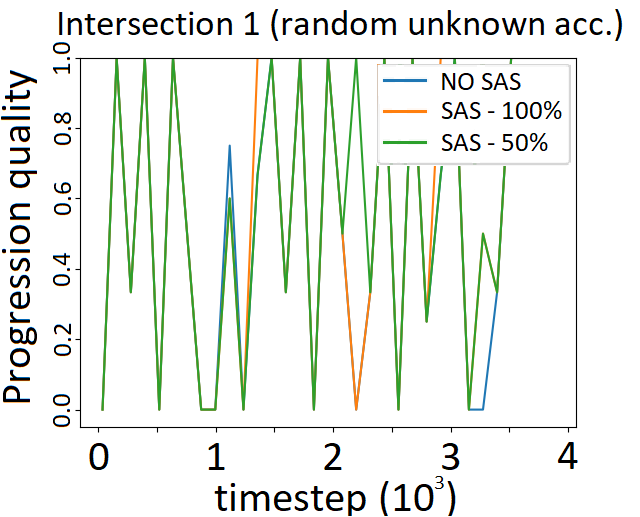}
		\caption{Low demand}
		\label{comPCa}
	\end{subfigure}
	\hfill
	\begin{subfigure}{0.32\linewidth}
	    \includegraphics[width=\linewidth]{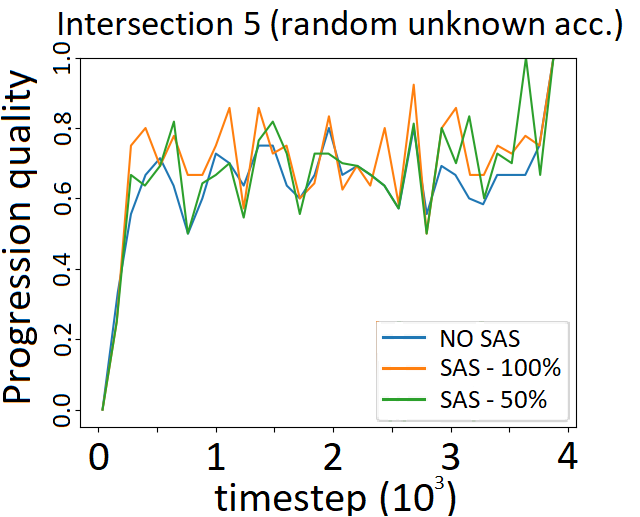}
		\caption{Medium demand}
		\label{comPCb}
  	\end{subfigure}
  	\hfill
	\begin{subfigure}{0.32\linewidth}
	    \includegraphics[width=\linewidth]{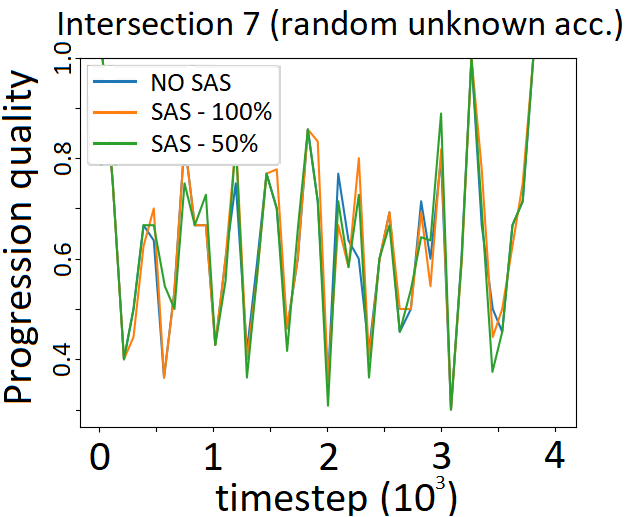}
    	\caption{High demand}
    	\label{comPCc}
  	\end{subfigure}
  	\hfill
	\caption{Progression quality for different SAS-equipped vehicle penetration levels.}
    \label{comPC}
\end{figure}

\subsubsection{Progression Quality}

progression quality comparison is presented in Fig. \ref{comPC}. Each graph contains outcomes for 0\%, 50\% and 100\% SAS-vehicle penetration. All three tested acceleration options are compiled by rows in the following order (starting from the top): known fixed, known random and unknown random accelerations. 

According to the results (Fig. \ref{comPCa}), progression quality for low traffic demand slightly benefits from the introduction of SAS-equipped vehicles. Being able to achieve 100\% POG more frequently implies that during some cycles we managed to eliminated delays completely. An important detail worth mentioning is that equipping 50\% of traffic participants with the Speed Advisory System results in almost the same improvement as making all of the vehicles controlled.

In the case of moderate traffic (Fig. \ref{comPCb}), the algorithm achieves more impressive results. Implementation of both 50\% and 100\% SAS penetration rates resulted in noticeable improvements in progression quality by at least 10\% and 8\% on average respectively. Moreover, in the case of known fixed accelerations and 100\% penetration rate the maximal PQ increase within one cycle was 20\%; and in the case of unknown random acceleration and 50\% penetration rate the growth reached 33\%. These results indicate that real-time prediction was able to provide accurate enough information to significantly improve traffic conditions at intersections in medium demand scenario.

Similarly to low demand, congested traffic (Fig. \ref{comPCc}) is not significantly affected by the Speed Advisory System in terms of progression quality. On average, the impact on the metric does not exceed 2\%, which correlates with the simple case results.

\section{Task distribution} \label{distrib}

It is important to highlight the responsibility distribution for each stage of the algorithm execution. 

First of all, as mentioned in the previous section, for each vehicle we need to estimate the time within a cycle when this vehicle arrives at the actuator $T_{est}^i$. This is expected to be done by the \textbf{infrastructure}, since the traffic data get stored at the road-side unit, i.e. computer, and no delay for data transmission is necessary. 

The next step is labeling vehicles with ``PASS" or ``WAIT" labels, which can be done by either the \textbf{infrastructure} or the \textbf{vehicle} itself. Choosing the first option, we face a challenge of distinguishing vehicles, finding the correct recipient for a particular message and transmitting the information to the vehicle in heavy traffic. On the other hand, by forcing \emph{vehicles} to conduct these calculations, we risk making a mistake in matching received data with the current vehicle's state due to delays and differences in timing. Further research is needed to address these issues.

The final stage of the procedure consists of computing the near-optimal speed trajectory, which is expected to be done by the Speed Advisory System installed at the \textbf{vehicle}.

\section{Conclusion}

The paper describes a real-time prediction algorithm that ties together two systems: Speed Advisory and actuated traffic light. The primary objective of the algorithm is determining the passing capability of a vehicles and assigning ``PASS" or ``WAIT" labels based on the collected traffic data and estimated downstream traffic state. 
The obtained information can be used in driver-assistance systems for building near-optimal speed trajectories to minimize fuel consumption. In addition, the algorithm provides an estimation of the current green phase duration for further computations.

Simulations demonstrate impressive results: at least 89\% prediction accuracy in the worst case and over 95\% correct estimations on average. The algorithm outperformed a statistically-based approach for both reliability levels of 0.8 and 0.1 for all traffic demands, SAS-vehicle penetration levels and acceleration scenarios. Integrating our algorithm into the procedure from \cite{15} can potentially extend the implementation of the method to primary actuated roads.

Significant fuel consumption reduction for both SAS-equipped (up to 30\%, 20\% and 7\% for low, medium and high demands respectively) and ordinary (up to 14\%) vehicles is another important outcome. Presence of vehicles with driver-assistance system forced ordinary cars to adjust their speed profiles and save up fuel. In addition, progression quality and phase utilization were improved for most of the scenarios.

Several issues have to be addressed in further studies. In this paper we did not consider queues and other forms of congestion, which can potentially have an impact on prediction accuracy. Switching from near-optimal speed trajectories to the car-following model compromises the prediction and can result in mismatches and errors.
Moreover, the algorithm has been tested only in SUMO, which does not fully reflect real world conditions. 

The algorithm might be viewed as a step towards a potentially high-impact system: a comprehensive intersection infrastructure that incorporates various road sensors, driver-assistance compatible software, hazardous behavior prevention and safety regulation. 

\bibliographystyle{unsrt}
\bibliography{citing}

\begin{thebibliography}{10}

\bibitem{1}
N.~Wan, A.~Vahidi, and A.~Luckow.
\newblock ``{O}ptimal speed advisory for connected vehicles in arterial roads
  and the impact on mixed traffic".
\newblock {\em Transportation Research Part C: Emerging Technologies},
  69:548--563, 2016.

\bibitem{2}
E.~Koukoumidis, L.-S. Peh, and M.~R. Martonosi.
\newblock ``{S}ignal{G}uru: leveraging mobile phones for collaborative traffic
  signal schedule advisory".
\newblock In {\em Proceedings of the 9th international conference on Mobile
  systems, applications, and services}, pages 127--140. ACM, 2011.

\bibitem{3}
B.~Asadi and A.~Vahidi.
\newblock ``{P}redictive cruise control: Utilizing upcoming traffic signal
  information for improving fuel economy and reducing trip time''.
\newblock {\em IEEE Transactions on Control Systems Technology},
  19(3):707--714, 2010.

\bibitem{4}
C.~Wang and S.~Jiang.
\newblock ``{T}raffic signal phases' estimation by floating car data.
\newblock In {\em 2012 12th International Conference on ITS
  Telecommunications}, pages 568--573. IEEE, 2012.

\bibitem{5}
V.~Protschky, C.~Ruhhammer, and S.~Feit.
\newblock ``{L}earning traffic light parameters with floating car data.
\newblock In {\em 2015 IEEE 18th International Conference on Intelligent
  Transportation Systems}, pages 2438--2443. IEEE, 2015.

\bibitem{6}
S.~A. Fayazi, A.~Vahidi, G.~Mahler, and A.~Winckler.
\newblock ``{T}raffic signal phase and timing estimation from low-frequency
  transit bus data".
\newblock {\em IEEE Transactions on Intelligent Transportation Systems},
  16(1):19--28, 2014.

\bibitem{7}
S.~A. Fayazi and A.~Vahidi.
\newblock ``{C}rowdsourcing phase and timing of pre-timed traffic signals in
  the presence of queues: algorithms and back-end system architecture".
\newblock {\em IEEE Transactions on Intelligent Transportation Systems},
  17(3):870--881, 2015.

\bibitem{8}
V.~Protschky, S.~Feit, and C.~Linnhoff-Popien.
\newblock ``{E}xtensive traffic light prediction under real-world conditions".
\newblock In {\em 2014 IEEE 80th Vehicular Technology Conference
  (VTC2014-Fall)}, pages 1--5. IEEE, 2014.

\bibitem{9}
V.~Protschky, K.~Wiesner, and S.~Feit.
\newblock ``{A}daptive traffic light prediction via {K}alman filtering".
\newblock In {\em 2014 IEEE Intelligent Vehicles Symposium Proceedings}, pages
  151--157. IEEE, 2014.

\bibitem{10}
S.~Ibrahim, D.~Kalathil, R.~O. Sanchez, and P.~Varaiya.
\newblock ``{E}stimating phase duration for {SP}a{T} messages.
\newblock {\em IEEE Transactions on Intelligent Transportation Systems}, 2018.

\bibitem{11}
J.~Sun and L.~Zhang.
\newblock ``{V}ehicle actuation based short-term traffic flow prediction model
  for signalized intersections".
\newblock {\em Journal of Central South University}, 19(1):287--298, 2012.

\bibitem{12}
S.~Chen and D.~J. Sun.
\newblock ``{A}n improved adaptive signal control method for isolated
  signalized intersection based on dynamic programming".
\newblock {\em IEEE Intelligent Transportation Systems Magazine}, 8(4):4--14,
  2016.

\bibitem{13}
S.~Coogan, C.~Flores, and P.~Varaiya.
\newblock ``{T}raffic predictive control from low-rank structure".
\newblock {\em Transportation Research Part B: Methodological}, 97:1--22, 2017.

\bibitem{14}
C.~M. Day, D.~M. Bullock, H.~Li, S.~M. Remias, A.~M. Hainen, R.~S. Freije,
  A.~L. Stevens, J.~R. Sturdevant, and T.~M. Brennan.
\newblock ``{P}erformance measures for traffic signal systems: An
  outcome-oriented approach".
\newblock Technical report, 2014.

\bibitem{15}
C.~Sun, X.~Shen, and S.~Moura.
\newblock ``{R}obust optimal eco-driving control with uncertain traffic signal
  timing".
\newblock In {\em 2018 Annual American Control Conference (ACC)}, pages
  5548--5553. IEEE, 2018.

\end{thebibliography}

%

\begin{IEEEbiography}[{\includegraphics[width=1in,height=1.25in,clip,keepaspectratio]{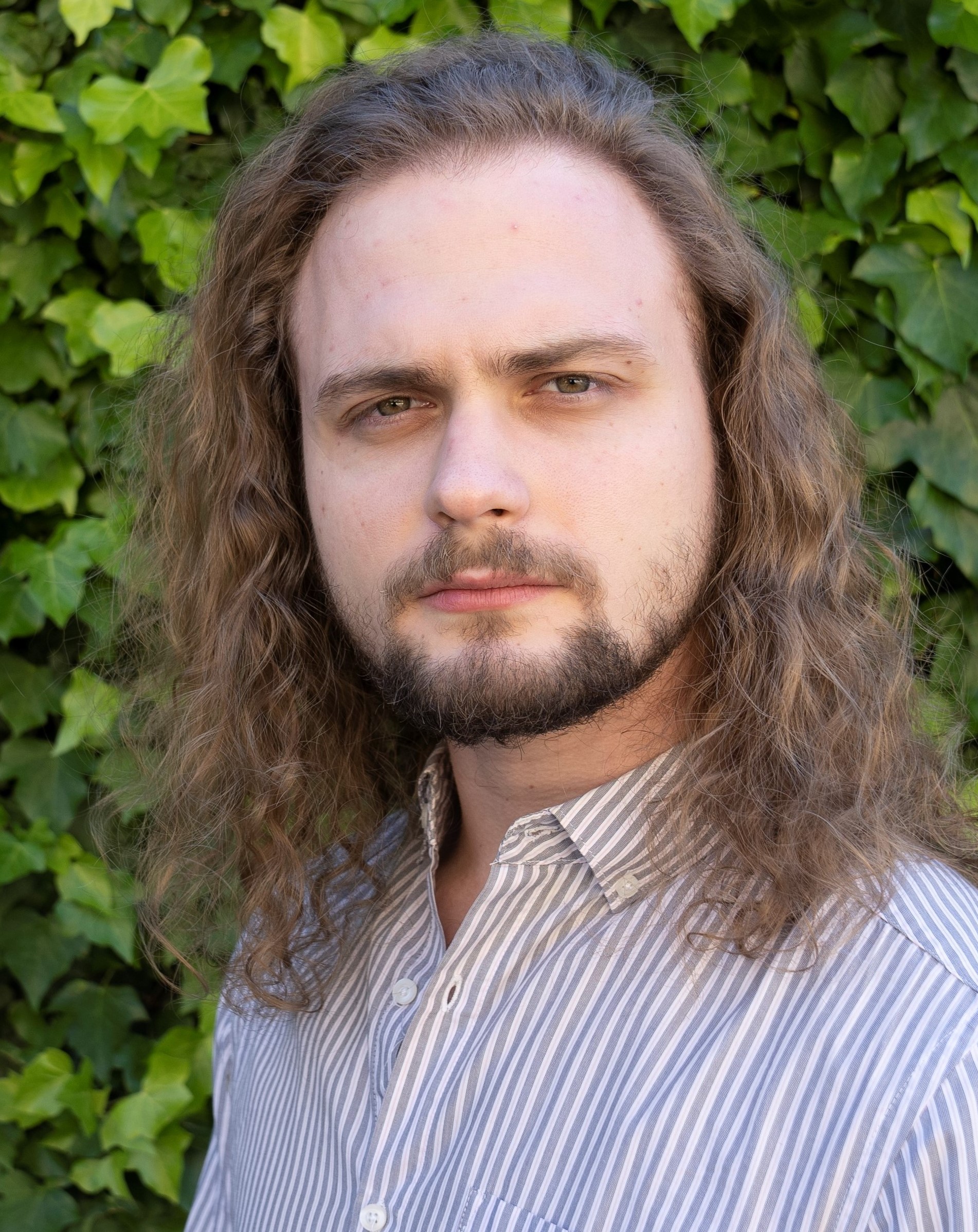}}]{Mikhail Burov}
received the Bachelor's degree in
control and applied mathematics from Moscow Institute of Physics and Technology, Russia, in 2017 and
the MS degree in electrical engineering and computer science from
University of California, Berkeley, CA, USA in 2019, where he is currently
pursuing the Ph.D. degree in electrical engineering and computer science.

From 2016 to 2017 he was an Undergraduate Research Assistant at Keldysh Institute of Applied Mathematics, Moscow, Russia. He is currently a Graduate Student Researcher with Pr. Murat Arcak's research group at University of California, Berkeley. His research interests include control systems, system optimization, autonomous driving and related applications.
\end{IEEEbiography}

\begin{IEEEbiography}[{\includegraphics[width=1in,height=1.25in,clip,keepaspectratio]{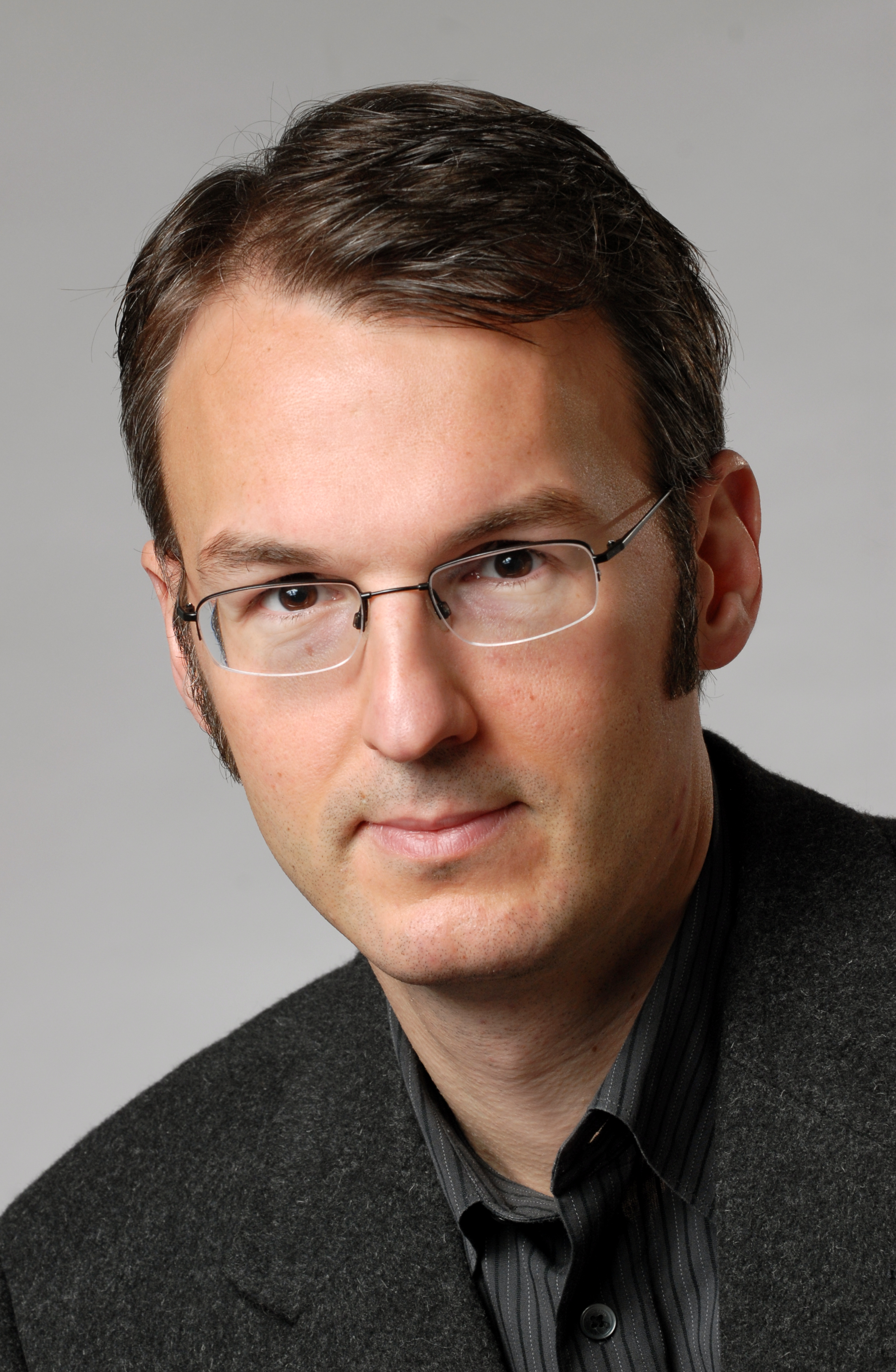}}]{Murat Arcak} is a professor at U.C. Berkeley in the Electrical Engineering and Computer Sciences Department, with a courtesy appointment in Mechanical Engineering.  He received the B.S. degree in Electrical Engineering from the Bogazici University, Istanbul, Turkey (1996) and the M.S. and Ph.D. degrees from the University of California, Santa Barbara (1997 and 2000). His research is in dynamical systems and control theory with applications to synthetic biology, multi-agent systems, and transportation. Prior to joining Berkeley in 2008, he was a faculty member at the Rensselaer Polytechnic Institute. He received a CAREER Award from the National Science Foundation in 2003, the Donald P. Eckman Award from the American Automatic Control Council in 2006, the Control and Systems Theory Prize from the Society for Industrial and Applied Mathematics (SIAM) in 2007, and the Antonio Ruberti Young Researcher Prize from the IEEE Control Systems Society in 2014. He is a member of ACM and SIAM, and a fellow of IEEE and the International Federation of Automatic Control (IFAC).
\end{IEEEbiography}


\begin{IEEEbiography}[{\includegraphics[width=1in,height=1.25in,clip,keepaspectratio]{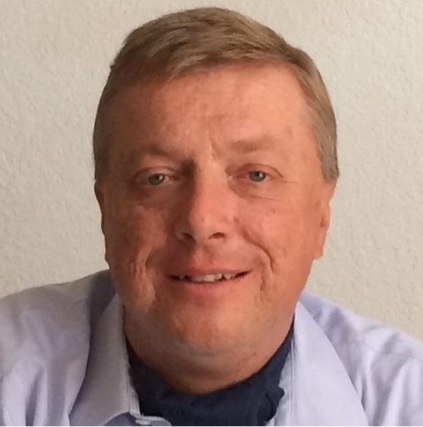}}]{Alexander Kurzhanskiy} received his M.S. in Applied Mathematics from the Lomonosov Moscow
State University in 1998 and Ph.D. in control and communication systems in 2007 from the
department of Electrical Engineering and Computer Science at UC Berkeley.

He joined California PATH as a Postdoc in 2008, and since 2011 works there as a Research
Engineer. He specializes in automatic control, large scale transportation data analysis, modeling
and simulation of traffic flows and individual vehicles, traffic control, vehicle automation and
intersection safety.
\end{IEEEbiography}




\end{document}